\documentclass[12pt]{article}

\usepackage{graphicx}
\usepackage{amsmath}
\usepackage{amssymb}
\usepackage{amsfonts}
\usepackage{graphicx}
\usepackage{mathrsfs}
\usepackage{dsfont}

\def\vhalf{\vphantom{\frac{1}{2}}}

\def\tpi{\widetilde{\pi}}

\def\tomega{\widetilde{\omega}}
\def\tX{\widetilde{X}}
\def\txx{\widetilde{x}}
\def\tg{\widetilde{g}}

\def\tphi{\widetilde{\phi}}
\def\tchi{\widetilde{\chi}}
\def\tpsi{\widetilde{\psi}}

\def\tone{\widetilde{1}}

\def\tc{\widetilde{c}}

\def\tm{\widetilde{m}}

\def\tR{\widetilde{R}}

\newcommand{\delr}{\raise.3ex\hbox{$\stackrel{\leftarrow}{\partial }$}}
\newcommand{\dell}{\raise.3ex\hbox{$\stackrel{\rightarrow}{\partial}$}}

\newcommand{\dr}{\raise.3ex\hbox{$\stackrel{\leftarrow}{\delta }$}}
\newcommand{\dl}{\raise.3ex\hbox{$\stackrel{\rightarrow}{\delta}$}}

 \def\balpha{{\overline{\alpha}}}
 \def\bbeta{{\overline{\beta}}}

 \def\bkappa{{\overline{\kappa}}}

 \def\bnu{{\overline{\nu}}}

 \def\btau{{\overline{\tau}}}

 \def\bomega{{\overline{\omega}}}
 
 \def\bone{{\overline{1}}}

\def\ext{\mathrm{ext}}

\def\min{\mathrm{min}}
\def\gh{\mathrm{gh}}
\def\bos{\mathrm{bos}}

\def\dem{\partial_{-}}
\def\dep{\partial_{+}}
\def\demm{\partial_{--}}
\def\depp{\partial_{++}}
\newcommand{\pip}[1]{\pi_{+ #1}}
\newcommand{\pim}[1]{\pi_{- #1}}
\newcommand{\apip}[1]{\pi_*^{+ #1}}
\newcommand{\apim}[1]{\pi_*^{- #1}}
\newcommand{\tpip}[1]{\tpi^{+ #1}}
\newcommand{\tpim}[1]{\tpi^{- #1}}
\newcommand{\atpip}[1]{\tpi^*_{+ #1}}
\newcommand{\atpim}[1]{\tpi^*_{- #1}}

\def\cM{{\cal{M}}}
\def\cD{{\cal{D}}}

\def\cO{{\cal{O}}}

\def\tcM{{\widetilde{\cal{M}}}}

\def\BRST{\mathrm{BRST}}
\def\BV{\mathrm{BV}}
\def\bos{\mathrm{bos}}

\catcode`\@=11
\def\marginnote#1{}

\def\titlepage{\@restonecolfalse\if@twocolumn\@restonecoltrue\onecolumn
     \else \newpage \fi \thispagestyle{empty}\c@page\z@
        \def\thefootnote{\fnsymbol{footnote}} }

\def\endtitlepage{\if@restonecol\twocolumn \else  \fi
        \def\thefootnote{\arabic{footnote}} \setcounter{footnote}{0}}

\catcode`@=12 \relax

\makeindex

\begin{document}
\topmargin -1.1cm
\begin{titlepage}
\begin{flushright}
arXiv:0809.4034
\end{flushright}
\vskip 2.5cm
\begin{center}
{\Large\bf Doubled Formalism, Complexification  }\\
\vspace{3mm} {\Large \bf and Topological $\sigma$-Models }\\
\vskip 1.5cm 
{\bf Vid Stojevic }\\
\vskip.6cm 
II. Institut f\"{u}r Theoretische Physik der Universit\"{a}t Hamburg,
\\
Luruper Chaussee 149, 22761 Hamburg, Germany \\ 
 {\tt vid.stojevic@desy.de} \\
 
\end{center}

\vskip .7cm

\begin{center}
{\bf Abstract}
\end{center}
\begin{quote}

We study a generalization of the Alexandrov-Kontsevich-Schwarz-Zaboronsky (AKSZ) formulation of the A- and B-models which involves a doubling of coordinates, and can be understood as a complexification of the Poisson $\sigma$-model underlying these.  In the flat space limit the construction contains models obtained by twisting an N=2 supersymmetric $\sigma$-model on Hull's doubled geometry. The curved space generalization involves a product of two diffeomorphic Calabi-Yau manifolds, and the $O(d,d)$ metric can be understood as a complexification of the CY metric. In addition, we consider solutions that can not  be obtained by twisting the above $\sigma$-model.  For these it is possible to interpolate between a model evaluated on holomorphic maps and one evaluated on constant maps by different choices of gauge fixing fermion.  Finally, we discuss some intriguing similarities between aspects of  the doubled formulation and topological M-theory, and a possible relation with results from the theory of Lie and Courant algebroids, where a doubled formulation plays a role in relating two- and three-dimensional topological theories.

\end{quote}
\end{titlepage}
\setcounter{footnote}{0} \setcounter{page}{0}
\setlength{\baselineskip}{.6cm}
\newpage
\begin{small}
\tableofcontents
\end{small}

\setcounter{equation}{0}


\section{Introduction}
\setcounter{equation}{0}
 
The Alexandrov-Kontsevich-Schwarz-Zaboronsky (AKSZ) formalism \cite{Alexandrov:1995kv} is an application of the Batalin-Vilkovisky (BV) quantization procedure  \cite{Batalin:1981jr, Batalin:1984jr} suited for the construction of topological quantum field theories. Starting from the AKSZ formulation of the A- and B-models, we  study an extension in which the number of coordinates is doubled, and which can be understood as an AKSZ model on a complexification of a Calabi-Yau manifold. The number of models that can be obtained from the doubled formulation is too large to consider in full generality, and in this paper we restrict the analysis to two quite general classes of solution to the master equation.  

For the first class, in the flat space limit, one recovers the doubled A- and B-models that are obtained by twisting an N=2 supersymmetric action on Hull's doubled geometry\footnote{Throughout the paper we will be referring to the scenario with all directions doubled.}  \cite{Hull:2004in, Hull:2005hk, Dabholkar:2005ve,Hull:2006va}, whose derivation is also a  new result of this paper. In curved space the solution generalizes to a product of two diffeomorphic Calabi-Yau manifolds whose metrics are locally the same, and  the role of the $O(d,d)$ metric of Hull's doubled formalism is played by a complexification of the Calabi-Yau metric. The precise meaning of complexification in this context is given in terms of a normal coordinate expansion, which is determined by the solution to the master equation. We also argue that the partition function of the doubled theory doesn't simply calculate the square of the A- and B-model partition function, but rather the squared norm.  This gives an indication of a possible relation with topological M-theory, since the Hitchin functional formulation of topological M-theory \cite{Dijkgraaf:2004te, Ooguri:2004zv} on a Calabi-Yau times a circle calculates the Wigner transform of $|Z_{A/B}|^2$.\footnote{The conjecture is verified at genus zero for the B-model;  at genus one the conjecture needs to be modified as described in  \cite{Pestun:2005rp}.}

The second type of solution is characterized by the property that one can interpolate between a model evaluated on holomorphic maps (A-type) and one evaluated on constant maps (B-type) simply by a choice of gauge-fixing fermion. A related observation is that in the flat space limit one can not obtain such models by twisting the N=2 action on Hull's doubled geometry. We consider a particular example in detail, and show that the A-type model is formulated in terms of the product of two manifolds, while the B-type model sees a complexification of a single manifold. By assuming an absence of a BV anomaly, which implies that A- and B-type models related by a gauge transformation are equivalent, we conjecture some possibilities as to what solutions to the master equation are possible when the target space is curved.  

Since the ideas of BV and AKSZ quantization are not widely familiar, we give a somewhat self contained review in section \ref{sec:BV_formalism}. In section \ref{sec:standard_constructions} we review the AKSZ constructions of the standard A- and B-models, and in section \ref{sec:doubled_picture} we discuss the details of the doubled formulation. In section \ref{conclusions} we  discuss a construction from the theory of Lie and Courant algebroids, in which a doubled formalism plays a role in obtaining a relation between topological string and membrane theories. An intermediate result that arises in this construction also offers some insight as to the relation between the two classes of solution to the master equation which we consider in this paper.

\section{The Batalin-Vilkovisky formalism and AKSZ}
\label{sec:BV_formalism}
\setcounter{equation}{0}

In the context of standard gauge theories, the BV quantization procedure provides a method for gauge fixing, by which we mean that it gives a prescription for obtaining an action without any local symmetries starting from a gauge theory action, and that it defines what the physical amplitudes are. At the same time, it turns out that the BV prescription naturally defines a supermanifold with a certain structure, which we will refer to as a BV geometry.  In Subsection \ref{subsec:BV_standard} we will explain how this geometry emerges from the gauge fixing perspective, but let us first define it.

Consider a supermanifold with an odd symplectic structure $( \cdot, \cdot)$ \cite{Schwarz:1992nx}. To obtain a BV manifold, we require that in addition it admits a volume element  $\sigma$ such that the Laplacian operator $\Delta$ defined by its action on a function $f$, $\Delta f := div K_f$, where $K_f :=  ( \cdot , f)$ obeys $\Delta^2 = 0$. Then one can show that given a function $F$ that obeys $\Delta F = 0$, the integral over a  Lagrangian submanifold\footnote{The definition of a Lagrangian submanifold in the context  of a BV manifold is the obvious extensions of the standard one: a submanifold  with half the maximal dimension such that the odd symplectic form restricted to it vanishes.}  $L$
\begin{equation}
\label{eq:odd_sym_form_integral}
\int_L \sigma F
\end{equation}
is invariant under deformations of $L$. If we identify the above with a path integral, then this theorem is tantamount to the independence of the quantum theory on the gauge choice.  To make contact with the standard expression for a path integral measure, we express $F$ in terms of a quantum BV action $S$, $F= \exp{\hbar^{-1} S}$. The equation $\Delta F = 0$ reads
\begin{equation}
\hbar \Delta S + (S,S) = 0 \ ,
\end{equation}
and is known as the quantum master equation. The part proportional to $\hbar$ can be understood as a quantum correction, so the classical BV action is required to obey the classical master equation
\begin{equation}
(S,S) = 0 \ .
\end{equation}
The geometric description of BV quantization was first given in \cite{Schwarz:1992nx} (see also \cite{Schwarz:2000ct}). In the terminology introduced in these papers a supermanifold with an odd symplectic structure is referred to as a P-manifold, while the BV manifold is referred to as an SP-manifold.

In quantum field theory, the BV manifold is an infinite dimensional space of fields, so $S$ is a functional and $\int_L \sigma \exp{\hbar^{-1} S}$ is a functional integral. The  measure $\sigma$ can usually not be rigorously defined. An action that obeys the classical master equation is  taken as the starting point in the perturbative evaluation of the quantum field theory, and the partition function is defined using the Feynman diagram expansion, and only makes sense if the theory is renormalizable. The quantum action $S_q$ can then only be understood in terms of regulators,  since in the limit when the regulators are taken to zero $S_q$ is infinite. The quantum master equation must also be regularized: it turns out that the $\Delta$ operator is ill-defined for infinite dimensional manifolds, and must itself be regularized. 

Therefore, we see that a QP-manifold \cite{Alexandrov:1995kv}, that is, a P-manifold equipped with an odd vector field 
\begin{equation}
\label{eq:Q_structure}
\widehat{Q} := ( \cdot, S)
\end{equation}
that obeys $\widehat{Q}^2=0$ (which is equivalent to the existence of a function(al) $S$ obeying $(S,S)=0$),  is the supergeometry corresponding to a gauge theory before quantum corrections are taken into account. Because a renormalization procedure is not needed in the evaluation of the topological theories of the type that we address in this paper, one can understand their content already from the solution of  the classical master equation. This is equivalently stated as the fact that topological theories can only depend on $\hbar$ via topological terms.  In the AKSZ procedure, which we describe in detail in  Subsection \ref{subsec:AKSZ},  a QP-manifold is constructed from standard geometrical structures, and this is enough to define the full quantum theory.

\subsection{BV: The standard approach}
\label{subsec:BV_standard}

In this section we describe how the BV geometry arises in gauge fixing a classical gauge invariant action $S_0$. Let us take $S_0$ to depend on a set of fields $\phi^i(x)$, and we express the gauge symmetries as
\begin{equation}
\label{eq:gauge_transf}
\delta \phi^i = \varepsilon^A R_A^i \ ,
\end{equation}
using the shorthand deWitt notation, where repeated indices signify  \emph{both} summation and integration and which will be used throughout this section. The gauge transformations are  labeled by the capital letter index, $\varepsilon^A$ stand for a set of transformation parameters depending on $x$, and written out in full the above expression reads:  
\begin{equation}
\varepsilon^A R_A^i  \equiv \int dz \varepsilon^A(z)  R_A^i(x-z) \ .
\end{equation}
Then gauge invariance of the action is expressed as
\begin{equation}
\label{eq:inv_of_action}
\frac{\dr S_0}{ \delta \phi^i} \varepsilon^A R_A^i =0 \ 
\end{equation}
in the deWitt notation.

The first step in the gauge fixing procedure is to introduce ghost fields $c^A$, which have opposite parity to the gauge transformation parameters $\varepsilon^A$. These are grouped together with $\phi^i$ under a collective field,
\begin{equation}
\label{eq:collective_field}
\Phi^\alpha = \{ \phi^i, c^A \} \ .
\end{equation}
To obtain the supermanifold on which the BV structure will be defined, we introduce the fields $\Phi^*_\alpha$, having opposite statistics to $\Phi^\alpha$, referred to as \emph{antifields}.  The P-structure is defined as 
\begin{equation}
\label{eq:antibracket_def}
(A, B) :=   \frac{ \dr A} { \delta   \Phi^\alpha } \frac{\dl B}{ \delta \Phi^*_\alpha} - 
\frac{\dr A}{ \delta \Phi^*_\alpha }  \frac{ \dl B}{ \delta   \Phi^\alpha }  \  .
\end{equation}
acting on two objects $A$ and $B$ that depend on $\Phi$ and $\Phi^*$.  

The Q-structure is defined by extending the classical action with terms depending on ghosts and antifields, as
\begin{equation}
\label{eq:min_solution}
S_\min = S_0 + \phi^*_i c^A R^i_A + \cdots  \ .
\end{equation}
For reasons that will become clear shortly, $S_\min$ is called the minimal solution. The dots are completed by requiring $S_\min$ to be a solution to the master equation
\begin{equation}
\label{eq:master_equation}
(S_\min, S_\min) = 0 \ ,
\end{equation}
For a gauge algebra that is not reducible and closes on-shell the solution to the master equation is given by
\begin{equation}
\label{eq:closed_alg_BV_action}
S_{\min} = S_0 + \phi^*_i c^A R_A^i + c^*_C N^C_{ \ AB} c^A c^B \ ,
\end{equation}
where $N^C_{ \ AB}$  are the (possibly field dependent) structure functions of the algebra. The master equation reads
\begin{align}
\frac{1}{2} (S_\min ,S_\min ) =  & \frac{ \dr S_0 }{\delta \phi^i} c^A R^i_A   + \phi^*_i \left[ c^A \frac{ \dr  R^i_A}{\delta \phi^k} c^B R^k_B + (-1)^{\epsilon_{\phi_i} \epsilon_{c^F}} R^i_F N^F_{ \ AB} c^A c^B \right] \\ \nonumber
& + c^*_D \left[2 N^D_{ \ AF} c^A N^F_{ \ GH} c^G c^H  - c^B c^C \frac{\dr N^D_{BC} }{\delta \phi^k} c^A R^k_A \right] = 0 \ ,
\end{align}
where $\epsilon \in \mathbb{Z}_2$ is zero when the field in its subscript is bosonic and one when it is fermionic. The term independent of antifields expresses the invariance of the action, the term proportional to $\phi^*$ the closure of the algebra, while the term proportional to $c^*$ is related to the Jacobi identity. For gauge algebras that close only off shell (open algebras), terms non-linear in the antifields are needed to obtain a solution to the master equation.

The gauge fixing step consists of picking a Lagrangian submanifold of the PQ-manifold. The obvious choice is given by setting all the antifields $\Phi^*$ to zero. This just leaves the standard gauge invariant action, which is clearly not a good starting point for defining the quantum theory.  So we seek a deformation away from this choice, by performing a canonical transformation, that is, a transformation $\Phi \rightarrow \Phi^{'}$, $\Phi^* \rightarrow \Phi^{*'}$,  that preserves the antibracket (\ref{eq:antibracket_def}). It can be shown that such a transformation is generated by a fermionic function $F(\Phi, \Phi^{*'})$ of fields and antifields as:
\begin{equation}
\label{eq:canonical_transf}
\Phi^{\alpha '}  = \frac{\delta F(\Phi, \Phi^{*'}) }{\delta \Phi^{*'}_\alpha} \ \ \ ,  \ \ \ \Phi^{*}_\alpha =  \frac{\delta F(\Phi, \Phi^{*'}) }{\delta \Phi^{A} }  \ .
\end{equation}
It turns out that for purposes of gauge fixing it is sufficient to consider a less general set of transformations when $F$ is of the form
\begin{equation}
F(\Phi, \Phi^{*'}) = \Phi^\alpha  \Phi^{*'}_\alpha + \Psi(\Phi, \Phi^{*'}) \ ,
\end{equation}
where $\Psi$ is referred to as the \emph{gauge fixing fermion}.  Furthermore, it is sufficient to let $\Psi$ depend only on fields, so that the canonical transformation acts only on antifields. In this case they are simply
\begin{equation}
\label{eq:gauge_fixing}
\Phi^*_\alpha \rightarrow \Phi^*_\alpha + (\Psi(\Phi), \Phi^*_\alpha ) \ \ \ ,  \ \ \ \Phi^\alpha \rightarrow \Phi^\alpha  \ .
\end{equation}
The reason why this is sufficient is that the above transformation deforms the classical action by terms coming from the antifields dependent part of the extended action, but doesn't generate field redefinitions, i.e. $\Phi^{\alpha'} = \Phi^{\alpha}$. In order to perform a field redefinition and  preserve the canonical form of the antibracket (\ref{eq:antibracket_def}), it is necessary to generate these via (\ref{eq:canonical_transf}). 

The minimal solution has a global $U(1)$  "ghosts" symmetry, where the $U(1)$ charges, referred to as ghost numbers, are conventionally assigned as
\begin{equation}
\gh (\phi^i) = 0 \ \ \ , \ \ \ \gh (c^A) = 1 \ \ \ ,  \ \ \ \gh (\Phi^*_\alpha) = - \gh (\Phi^\alpha) - 1 \ .
\end{equation}
In order for the canonical transformation to preserve ghost number it's necessary that $\gh(F) = \gh (\Psi) = -1$.  We need a $\Psi$ independent of antifields, but such a fermion can't be constructed from the fields in the minimal solution, since these all have positive ghost number. To cure this auxiliary field pairs $b^A$ and $\lambda^A$ are introduced, with $\gh (b^A) = -1$ and $\gh (\lambda^A) = 0$, which can then be used to construct an appropriate $\Psi$. The extended action with the auxiliary fields (the non-minimal solution) reads
\begin{equation}
\label{eq:me_with_aux}
S_{\ext} = S_\min + b^*_A \lambda^A \ .
\end{equation}
After performing an appropriate canonical transformation one obtains an action that, after setting the antifields to zero, does have a well defined propagator. In fact, what is meant by "appropriate" is simply that a propagator exist, as there is no unique recipe for choosing the fermion; there are many gauge choices that do the job. The crucial ingredient that guarantees the independence of the theory on this choice, in the language of BV geometry, is the theorem mentioned in the context of (\ref{eq:odd_sym_form_integral}). 

The solution to the master equation defines a nilpotent operator, $\delta_{\BV}$:
\begin{equation}
\label{eq:BVoperator}
\delta_{\BV} A : = (A, S_\ext ) \ .
\end{equation}
$\delta_\BV \Phi^\alpha$ is independent of antifields only if the solution to the master equation is linear in the antifields (\ref{eq:closed_alg_BV_action}), in which case  it corresponds to the standard BRST transformations. It is conventional to define the BRST operator in the general case as
\begin{equation}
\delta_{\BRST} \Phi^\alpha =  \delta_{\BV}  \Phi^\alpha |_{\Phi^*_\alpha = 0} \ ,
\end{equation}
however, this operator then only nilpotent up to the equations of motion whenever $S_\ext$ has terms nonlinear in antifields. At the linear level antifields act as sources for the BRST transformations. In the quantum theory, the Ward identities derived from the BRST transformations are expressed using the partition function perturbed by the antifield sources, namely $\int [ d \Phi ] \exp{(i S_\ext) } $. This is essentially (\ref{eq:odd_sym_form_integral}) with the antifields treated as background fields in the path integral. In this paper we will often write down extended actions as a convenient way of expressing the BRST transformations together with the classical action.

The transformation  (\ref{eq:BVoperator}) also defines physical observables. Without any restrictions on the observables the theory would be non-unitary, since the ghost fields don't obey the spin-statistics theorem.  The natural restrictions is to require observables to  obey $\delta_{\BV} \cO   = 0$, because then one can show that their expectation values are independent of the gauge choice (this is a simple extension of the theorem that guarantees the gauge independence of the partition function).  However, any observable of the form $ \cO = \delta_{\BV} F$ can be considered trivial since it is automatically closed, and it follows that physically distinct observables are classified by the  cohomology of $\delta_{\BV}$. In order to respect the classical limit, a further restriction is to require observables to have ghost number zero.

\subsection{BV: The AKSZ approach}
\label{subsec:AKSZ}

In the AKSZ approach \cite{Alexandrov:1995kv}  the idea is to construct a PQ-manifold starting from a standard geometric structure, rather than from a classical gauge theory.  In very general terms, the idea is to construct two finite dimensional PQ-manifolds, one intended to be the target space and the other the base space of the $\sigma$-model, and then define a quantum field theory using a canonical PQ structure defined on the space of embeddings of the base space into the target space. The reason why it is sufficient to consider a PQ geometry to define the theory, rather than the full BV geometry, is because the theories constructed in this geometric manner turn out to be topological.

The PQ-manifolds we'll consider in this paper will be cotangent and tangent bundles over a Poisson manifold, with parity reversed fibers. Let us take the target space base manifold as $\cM$, then the parity reversed cotangent bundle is denoted as $\Pi T^* \cM$, where in general $\Pi$ will stand for parity reversal. The PQ structure on   $\Pi T^* \cM$ is defined as follows. Take  $X^i$ to be coordinates on $\cM$ and $\pi_i$ the odd coordinates on the fibers, the odd symplectic bracket is given by
\begin{equation}
(A, B) = \frac{\delr A}{\partial X^i} \frac{\dell B}{\partial \pi_i} -   \frac{\delr A}{\partial \pi_i} \frac{\dell B}{\partial X^i}  \ ,
\end{equation}
and the Q-structure is obtained from
\begin{equation}
S = \frac{1}{2} P^{ij}(X) \pi_i \pi_j \ ,
\end{equation}
where the master equation is satisfied provided that $P^{ij}$ defines a Poisson structure on $\cM$, i.e. $P^{[ ij}_{ \ \ ,k} P^{m ]k} = 0$. 

 Since we are considering string theories, the base space will be a parity reversed tangent bundle over a two-dimensional worldvolume,  $\Pi T \Sigma$.  There is a canonical PQ structure on the space of maps from $\Pi T \Sigma$ to $\Pi T^* \cM$.  Take the local coordinates on $\Pi T \Sigma$ to be 
\begin{equation}
\mu := \{ z^{+}, z^-, \theta^+ , \theta^- \} \  ,
\end{equation}
where $z^+$ are holomorphic and $z^-$ anti-holomorphic worldsheet coordinates, and $\theta^\pm$ are the fermionic fiber coordinates. The fields $X(\mu)$ and $\pi(\mu)$,  are referred to as \emph{de Rham superfields}. We note that $X(\mu)$ is a map  $\Pi T \Sigma \rightarrow \cM$, while $\pi(\mu)$ is a section of $ \Pi T^* \Sigma \otimes X^* ( \Pi T \cM)$. The P-structure is given by 
\begin{equation}
\label{eq:A_model_odd_symp_struct}
(A, B) :=   \int d^2 \mu \left( \frac{ \dr A} { \delta   X^i(\mu)  } \frac{\dl B}{ \delta \pi_i(\mu) } - 
\frac{\dr A}{ \delta \pi_i(\mu)}  \frac{ \dl B}{ \delta   X^i(\mu) }  \right)  \ .
\end{equation}
The Q structure is defined from the action
\begin{equation}
\label{eq:Poisson_action}
S = \int d^2 \mu \frac{1}{2} P^{ij}(X) \pi_i (\mu) \pi_j (\mu)  \ ,
\end{equation}
where
\begin{equation}
d^2 \mu = d z^+ d z^-  d \theta^+ d \theta^- \ .
\end{equation}
The ghost numbers are assigned as
\begin{equation}
\label{eq:A_model_gh_no}
\gh(X) = 0 \ \ \  , \ \ \  \gh (\pi) = -1 \  \ \  \mathrm{and} \ \ \  \gh(\theta^{\pm}) = 1 \ ,
\end{equation}
so that $\gh(d^2 \mu ) = -2$, and $\gh (S) = 0$.

The action (\ref{eq:Poisson_action}) is actually that of the standard Poisson $\sigma$-model  \cite{Cattaneo:1999fm}, which can be seen by expanding in terms of the worldsheet superspace coordinates,
\begin{align}
\label{eq:superfield_exp}
& X^k = \phi^k + \theta^- \pi_*^{+k} + \theta^+ \pi_*^{-k} - \theta^+ \theta^- \chi^k_* \ , \\ \nonumber
& \pi_k = \chi_k + \theta^- \pi_{-k} + \theta^+ \pi_{+k} + \theta^+ \theta^- \phi^*_k \ ,
\end{align}
where an initial choice of Lagrangian submanifold has been made, i.e. we have chosen which component fields are to be treated as fields and which as antifields.  This should be contrasted with the gauge fixing in the previous subsection, where it was clear, by construction, what the fields and the antifields were. Here we need to make a choice, and there is no guarantee that different choices will give the same theory. Examples where they do not do so have been given in  \cite{Stojevic:2008qy, Rogers:2005zk}, and a further example will be given in  the next subsection, where we show that the B-model can essentially be obtained starting from the same AKSZ action as the A-model, (\ref{eq:Poisson_action}), by making a distinct choice of  Lagrangian submanifold. The antibracket is expressed in components as:
\begin{align}
\label{eq:poisson_antibracket}
(A,B) = &  \int d^2 z \left( \frac{ \dr A} { \delta   \phi^i } \frac{\dl B}{ \delta \phi^*_i} - \frac{\dr A}{ \delta \phi^*_i}\frac{ \dl B}{ \delta   \phi^i } 
 - \frac{ \dr A} { \delta   \chi_i } \frac{\dl B}{ \delta \chi^i_*} +  \frac{\dr A}{ \delta \chi^i_*} \frac{ \dl B} { \delta   \chi_i } \right.  \\ \nonumber
& \left. + \frac{ \dr A} { \delta   \pi_{- i} } \frac{\dl B}{ \delta \pi_*^{- i} } - \frac{\dr A}{ \delta \pi_*^{- i} } \frac{ \dl B} { \delta   \pi_{- i} } 
 - \frac{ \dr A} { \delta   \pi_{+ i} } \frac{\dl B}{ \delta \pi_*^{+ i} }  + \frac{\dr A}{ \delta \pi_*^{+ i} } \frac{ \dl B} { \delta   \pi_{+ i} }  \right) \ .
\end{align}
Integrating over the fermionic coordinates in (\ref{eq:Poisson_action}) we obtain the extended action:
\begin{align}
\label{eq:poisson_action}
S =& \int d^2 z \left( \vphantom{\frac{1}{2}}  P^{ij} \pip{i} \pim{j} -  \pi_*^{+k}  P^{ij}_{ \ \ ,k} \chi_i \pip{j}  + \pi_*^{-k}  P^{ij}_{ \ \  ,k} \chi_i \pim{j}   \right.\\ \nonumber
& \left.  \vphantom{\frac{1}{2}}  + \phi^*_j P^{ji} \chi_i - \frac{1}{2} \chi_*^l P^{ij}_{ \ \  ,l} \chi_i \chi_j + \frac{1}{2} \pi_*^{+l} \pi_*^{-m}  P^{ij}_{ \ \ , lm}  \chi_i \chi_j \right) \ ,
\end{align}
where  $P^{ij}_{ \ \ ,k} : = \frac{\partial P^{ij}}{\partial \phi^k}$.  In terms of the standard BV construction (see in particular (\ref{eq:closed_alg_BV_action})), the  first term corresponds to the classical action, $\chi_k$ are the ghost fields, the gauge symmetry transformations can be read off from the terms linear in $\pi_*$ and $\phi^*$, and the structure constants of the algebra from the term proportional to $\chi_*$. The term quadratic in the antifields is present due to the algebra of these symmetries closing only on-shell.  

Finally, there is the following important distinction to the  gauge fixing procedure of the last subsection. There, in order to respect the classical limit, it was crucial that the fields of the classical action and the observables had ghost number zero. This requirement can be dropped for topological theories, which in particular means that it is not a priori necessary to introduce auxiliary pairs.

\section{The AKSZ construction of A- and B-models}
\label{sec:standard_constructions}
\setcounter{equation}{0}

One way to obtain the A-model  \cite{Alexandrov:1995kv} is to start with the Poisson $\sigma$-model of the last subsection  (\ref{eq:Poisson_action}), with $P^{ij}$ given by the inverse of the K\"{a}hler form, which we denote by $\omega$. The choice of Lagrangian submanifold in (\ref{eq:superfield_exp}) is in fact the correct one for the quantization of topological models related to deformation quantization  \cite{Cattaneo:1999fm, Cattaneo:2001ys, Cattaneo:2001bp}. In order to obtain the A-model one needs to choose a more "refined" Lagrangian submanifold by making use of an almost complex structure in the target space (in what follows we are taking the target space to be K\"{a}hler, and the structure to be actually complex, but this is not strictly necessary \cite{Witten:1988xj}):
\begin{align}
\label{eq:A_model_gauge_fixing}
\pi_{+ \alpha} & \rightarrow \psi^-_{* \alpha}   &  \pi_{+ \balpha} & \rightarrow   \pi_{+ \balpha}   
&  \pi_{ - \balpha} & \rightarrow \psi^+_{* \balpha}   & \pi_{- \alpha} &  \rightarrow \pi_{ - \alpha} \\ \nonumber
\pi^{+ \alpha}_*  & \rightarrow \psi_-^{\alpha}  &  \pi_*^{+ \balpha} & \rightarrow  \pi_*^{+ \balpha} 
& \pi_*^{- \balpha} & \rightarrow \psi_+^{\balpha}  &  \pi_*^{-\alpha} & \rightarrow \pi_*^{- \alpha} \ ,
\end{align}
referred to (anti)-holomorphic coordinates $X^\alpha$ ($X^\bbeta$). In addition, it is necessary to make an infinitesimal canonical transformation (\ref{eq:gauge_fixing}) away from this choice, generated by the fermion
\begin{equation}
\label{eq:A_model_fermion}
\Psi = - \int d^2 z  \omega_{\alpha \bbeta} \left( \dem \phi^\alpha \psi_+^\bbeta - \dep \phi^\bbeta \psi_-^\alpha \right) \ .
\end{equation}
After integrating out the auxiliary fields $\pi_{+\alpha}$ and $\pi_{- \bbeta}$ one obtains the BRST closed part of the standard A-model action (up to an overall factor of $i$),
\begin{align}
\label{eq:A_model}
S  = & \int d^2 z \left[ \vphantom{\frac{1}{2}}  i g_{\alpha \bbeta} \dep \phi^\alpha \dem \phi^\bbeta + \chi_{\nu}(\dep \psi_-^\nu + \Gamma^\nu_{ \ \alpha \beta} \dep \phi^\alpha \psi_-^\beta) \right. \\ \nonumber 
& - \chi_{\bnu}(\dem \psi_+^\bnu + \Gamma^\bnu_{ \ \balpha \bbeta} \dem \phi^\balpha \psi_+^\bbeta)  - i R^{\alpha \bbeta}_{ \ \ \kappa \bnu} \psi_-^\kappa \psi_+^\bnu \chi_\alpha \chi_\bbeta  \\ \nonumber
& +i \phi^*_\nu g^{\nu \bkappa} \chi_\bkappa - i \phi^*_\bnu g^{\bnu \kappa} \chi_\kappa + \psi_{*\beta}^- ( \dem \phi^\beta  + i  \Gamma^{\beta \bbeta}_{ \ \ \alpha} \psi^\alpha_- \chi_{\bbeta}) \\ \nonumber
& - \psi_{*\bbeta}^+ ( \dep \phi^\bbeta  - i  \Gamma^{\bbeta \beta}_{ \ \ \balpha} \psi^\balpha_+ \chi_{\beta}) + i \chi_*^\alpha \Gamma^{\nu \bkappa}_{ \ \ \alpha} \chi_\nu \chi_\bkappa + i \chi_*^\balpha \Gamma^{\bkappa \nu}_{ \ \  \balpha} \chi_\nu \chi_\bkappa \\ \nonumber
& + i g^{\alpha \bbeta} \psi_{* \alpha}^- \psi_{* \bbeta}^+  \left. \vphantom{\frac{1}{2}} \right] \ ,
\end{align}
where the convention is $\omega_{\alpha \bbeta} =  i g_{\alpha \bbeta}$. To obtain the full A-model action a topological term, namely the pullback of the K\"{a}helr form to the worldsheet, must be added:
\begin{equation}
\label{eq:A_model_top_term}
S_\omega = \frac{1}{2} \int d^2 \mu \omega_{ij} DX^i DX^j \equiv \frac{1}{2} \int d^2 z \omega_{ij} \dep \phi^i \dem \phi^j ,
\end{equation}
where
\begin{equation}
D := \theta^+ \dep +  \theta^- \dem \  \ ,
\end{equation}
and obeys $D^2=0$. This is a topological term due to $d\omega=0$, and  $S^{'} =   S + S_\omega$ still satisfies the master equation, and because it can be written directly in superfields, there is no problem in adhering fully to the AKSZ procedure and including it from the start (\ref{eq:Poisson_action}). A closed $b$-field can be introduced in the same manner: $S_{\omega} \rightarrow S_{\omega + b}$.

According to the fixed point theorem \cite{Witten:1991zz}, the path integral only obtains contributions from maps for which the BRST transformation of the fermions vanish. For (\ref{eq:A_model}) this is satisfied by setting the fermions to zero, and for $\phi^\alpha$ evaluated on holomorphic maps:
\begin{equation}
\dem \phi^\alpha = \dep \phi^\bbeta = 0 \ .
\end{equation}
In  \cite{Marino:2004eq, Marino:2004uf} a different argument is given, which states that the model is evaluated on maps  for which the part of the action given by $\delta_{\BRST} \Psi$ is identically zero. From the AKSZ point of view the two arguments can easily be seen to be equivalent, since in the extended action both the BRST variation of the fermions and $\delta_{\BRST} \Psi$ are obtained by the canonical transformation of the part of the extended action quadratic in $\psi_*$.

When comparing the above derivation of the A-model with the one obtained from twisting, an apparent discrepancy is that in the former the $\chi$ fields have a downstairs index, whereas in the latter the index is upstairs. One obtains observables by associating $\chi^i$ with the differentials $d\phi^i$ on the target space, since $\delta_\BRST \phi^i = \chi^i$, $\delta_\BRST \chi^i = 0$, and the observables are therefore in correspondence with the de Rham cohomology of $\cM$. In  (\ref{eq:A_model}) it is $g^{ij} \chi_j$ that are associated with $dz^i$, and this is the reason why $\chi_i$ don't transform trivially, as can be gathered from the presence of terms linear in $\chi_*$.  To put the action in the standard A-model form, one needs to make the field redefinition $\chi^{i'} = \omega^{ij} \chi_j$, and as was discussed in \ref{subsec:BV_standard}, in order to preserve the antibracket, this must be done via a canonical transformation (\ref{eq:canonical_transf}). The appropriate generating fermion is
\begin{equation}
\label{stand_A_model_canonical_trans}
F = \int d^2 z \left[  \chi^{*'}_i \omega^{ij} \chi_j + \Phi^{*'}_A \Phi^A \right] \ ,
\end{equation}
where $\Phi^A$ stands for all the fields in the action. This transformation precisely eliminates the $\chi_*$ term in (\ref{eq:A_model}).

The A-model can also be obtained from an AKSZ action whose target space is $\Pi(T \cM \oplus T^* \cM )$ \cite{Ikeda:2007rn}
\begin{equation}
\label{eq:A_model2}
S = \int d^2 \mu \left[ \frac{1}{2} \omega_{ij} \tpi^i \tpi^j + \pi_i \tpi^i \right]  \ ,
\end{equation}
where $\tpi^i$ are taken to be the fiber coordinates on $\Pi T \cM$. In order to define a P-structure it is also necessary to introduce a set of fields $\tX_k$  to which $\tpi^k$ are the antifield partners. The antibracket is given by
\begin{align}
\label{eq:doubled_model_odd_symp_struct}
(A, B) :=   \int d^2 \mu \left( \frac{ \dr A} { \delta   X^i  } \frac{\dl B}{ \delta \pi_i } 
-  \frac{\dr A}{ \delta \pi_i}  \frac{ \dl B}{ \delta   X^i }  
+ \frac{ \dr A} { \delta   \tX_i  } \frac{\dl B}{ \delta \tpi^i } -  \frac{\dr A}{ \delta \tpi^i}  \frac{ \dl B}{ \delta   \tX_i }     \right)  \ ,
\end{align} 
the additional de Rham superfields $\tX(\mu)$ and $\tpi(\mu)$  are expanded as,
\begin{align}
\label{eq:doubled_expansion}
\tX_k & = \tphi_k + \theta^- \atpim{k}   + \theta^+ \atpip{k} - \theta^+ \theta^- \tchi^*_k  \\ \nonumber
\tpi^k  & = \tchi^k + \theta^- \tpip{k} + \theta^+ \tpim{k} + \theta^+ \theta^- \tphi^k_* \ ,
\end{align}
and the component BV bracket is given by the obvious extension of (\ref{eq:poisson_antibracket}) to the doubled fields. The action (\ref{eq:A_model2}) satisfies the master equation if $d \omega = 0$. We will also consider the following action from which the A-model can be derived, and which explicitly involves both the complex and the K\"{a}hler structures of $\cM$:
\begin{equation}
\label{eq:A_model_orig2}
S = \int d^2 \mu \left[ \omega_{ \alpha \bbeta} \tpi^\alpha \tpi^\bbeta + i \pi_\alpha \tpi^\alpha -i \pi_\bbeta \tpi^\bbeta \right]  \ .
\end{equation}
It also  satisfies the master equation provided that $d\omega = 0$.  As a somewhat heuristic argument for the equivalence of the above two actions with the formulation in terms of $\omega^{-1}$, we note that eliminating the $\tpi$ fields for either action gets us back to (\ref{eq:Poisson_action}) up to overall normalization. 

In \cite{Alexandrov:1995kv} the B-model is essentially obtained from the second of the above actions\footnote{It is also possible to consider the smaller bundle $\Pi( T^{(0,1)} \cM \oplus T^{*(1,0)} \cM)$ as the target space for the B-model \cite{Hofman:2002cw, Pestun:2006rj}.}, after setting the $\omega$ term to zero, 
\begin{equation}
\label{eq:AKSZ_B_model}
S = \int d^2 \mu (i \pi_\alpha \tpi^\alpha -i \pi_\bbeta \tpi^\bbeta ) \ .
\end{equation}
The notation we use doesn't follow \cite{Alexandrov:1995kv}, but is more in line with the doubled extension we propose in the next sections. In components (\ref{eq:AKSZ_B_model}) reads:
\begin{align}
S = &  \int d^2 z \left[ \vphantom{\frac{1}{2}} 
-i \pim{\alpha} \tpim{\alpha} +i \pip{\alpha} \tpip{\alpha} + i \pim{\bbeta} \tpim{\bbeta} -i \pip{\bbeta} \tpip{\bbeta}  -i \tphi^\alpha_* \chi_\alpha+i \tphi^\bbeta_* \chi_\bbeta +i \phi^*_\alpha \tchi^\alpha - i \phi^*_\bbeta \tchi^\beta  \vphantom{\frac{1}{2}} \right] \ .
\end{align}
After making the following choice for the Lagrangian submanifold,
\begin{align}
\label{eq:B_model_standard_gauge}
 \pip{\alpha} & \rightarrow \psi^-_{* \alpha}   & \pip{\bbeta} & \rightarrow \pip{\bbeta}    & \pim{\alpha} & \rightarrow \psi^+_{* \alpha} & \pim{\bbeta} & \rightarrow  \pim{\bbeta}   \\ \nonumber
 \tpip{\alpha} & \rightarrow \tpsi^{*\alpha}_-   & \tpip{\bbeta}  & \rightarrow \tpip{\bbeta}  &   \tpim{\alpha} & \rightarrow \tpsi^{* \alpha}_+  &  \tpim{\bbeta} & \rightarrow \tpim{\bbeta}   \ ,
\end{align}
together with the corresponding choices for $\tpi^*$ and $\pi_*$, and performing a canonical transformation generated by
\begin{equation}
\label{eq:stand_B_model_ferm}
\Psi = \int d^2 z \left[ \vhalf \omega_{\alpha \bbeta} \psi_+^\alpha \dem \phi^\bbeta + \omega_{\alpha \bbeta} \psi_-^\alpha \dep \phi^\bbeta + i \tpsi^+_\alpha \dep \phi^\alpha - i \tpsi^-_\alpha \dem \phi^\alpha \right]  \ ,
\end{equation}
one obtains the extended action
\begin{align}
\label{eq:stand_B_model}
S  = & \int d^2 z \left[ \vhalf \omega_{\alpha \bbeta} \dem \phi^\bbeta \dep \phi^\alpha + \omega_{\alpha \bbeta} \dep \phi^\bbeta \dem \phi^\alpha  \right.   \\ \nonumber 
& - i \omega_{\alpha \bbeta} \tchi^\bbeta ( \dem \psi^\alpha_+ + \dep \psi_-^\alpha )  - i \tchi^\alpha ( \dep \tpsi^+_\alpha - \dem \tpsi^-_\alpha)  \\ \nonumber 
& +  \psi^+_{* \alpha} \dep \phi^\alpha + \psi^-_{* \alpha} \dem \phi^\alpha  + i \omega_{\alpha \bbeta} ( -\tpsi^{* \alpha}_+ \dem \phi^\bbeta + \tpsi^{* \alpha}_- \dep \phi^\bbeta )  \\ \nonumber
& \left.    -i \tphi^\alpha_* \chi_\alpha+i \tphi^\bbeta_* \chi_\bbeta +i \phi^*_\alpha \tchi^\alpha - i \phi^*_\bbeta \tchi^\beta -i \psi^+_{* \alpha} \tpsi^{* \alpha}_+  + i \psi^-_{*\alpha} \tpsi^{* \alpha}_-  \vphantom{\frac{1}{2}} \right]  \ .
\end{align}
The auxiliary fields have been eliminated (in this case the equations of motion simply set them to zero). By setting $\tpsi = \tpsi^* = \tphi_* = 0$, as well as $\phi^*_\alpha = 0$ we obtain the extended action of the $\delta_{\BRST}$  exact part of the B-model in flat space.  To obtain a version valid in curved space, one would have to gauge fix using a covariant version of (\ref{eq:stand_B_model_ferm}). The term analogous to (\ref{eq:A_model_top_term}) in the A-model, that must be added in order to obtain the full B-model extended action, now involves the fermions \cite{Witten:1991zz}:
\begin{equation}
\label{eq:B_standard_top_term}
\int d^2 z \chi_\alpha  ( \dep \psi^\alpha_- - \dem \psi^\alpha_+ ) \ .
\end{equation}
One can easily check that this is BRST invariant, even though it is not topological in the same sense as  (\ref{eq:A_model_top_term}). The fixed point theorem implies that the B-model is evaluated on constant maps,
\begin{equation}
\dep \phi^ \alpha = \dep \phi^\alpha = 0 \ ,
\end{equation}
and with the fermions set to zero. Observables are given by $(0,p)$-forms with values in $\wedge^q T^{(1,0)} \cM$, that is, objects of the form
\begin{equation}
\label{eq:stand_B_model_obs}
W^{\alpha_1 \cdots \alpha_q}_{ \ \ \  \ \ \ \ \bbeta_1 \cdots \bbeta_p} \chi_{\alpha_1} \cdots \chi_{\alpha_q} \tchi^{\bbeta_1} \cdots \tchi^{\bbeta_p}  \ ,
\end{equation}
where the association is 
\begin{equation}
\chi_\alpha \approx \frac{\partial}{\partial \phi^\alpha}  \ \ \  ,  \ \ \  \tchi^\bbeta \approx d\phi^\bbeta  \ .
\end{equation}

This particular derivation involves treating \emph{all} of the coordinates $\phi^\alpha$, $\phi^\bbeta$, $\tphi_\alpha$ and $\tphi_\bbeta$ as independent coverings of $\cM$ during the gauge fixing procedure, and correspondingly as independent fields in the $\sigma$-model, and only imposing a reality condition at the end. This should be contrasted with the doubled interpretation we propose in the following section, when the tilde and non-tilde fields are treated symmetrically during the gauge fixing, but $\phi^\balpha$ / $\phi^\alpha$ (and $\tphi_\alpha$ /  $\tphi_\balpha$) are \emph{not} taken to be independent. 

The B-model can also be derived starting from the same AKSZ action as the A-model (\ref{eq:Poisson_action}). We make the choice of Lagrangian submanifold (\ref{eq:B_model_standard_gauge}), excluding the tilde fields,  and then preform the canonical transformation generated by
\begin{align}
\label{eq:B_special_fermion}
\Psi  =  \int d^2 z \omega_{\alpha \bbeta}  \left[ \vhalf  \psi^\alpha_- \dep \phi^\bbeta + \psi_+^\alpha \dem \phi^\bbeta \right] .
\end{align}
Having also performed the transformation (\ref{stand_A_model_canonical_trans}) one obtains the action
\begin{align}
\label{eq:B_model_formulation2}
S = &  \int d^2 z \left[ i \vhalf  \pim{\bbeta} \dep \phi^\bbeta + i \pip{\bbeta} \dem \phi^\bbeta + \omega_{\alpha \bbeta} \psi^\alpha_- \dep \chi^\bbeta + \omega_{\alpha \bbeta} \psi^\alpha_+ \dem \chi^\bbeta \right. \\ \nonumber
 & \left. + \phi^*_\alpha \chi^\alpha + \phi^*_\bbeta \chi^\bbeta + \omega ^{\alpha \bbeta} \psi^{-}_{ \alpha*} \pim{\bbeta} + \omega^{\bbeta \alpha} \psi^+_{* \alpha} \pip{\bbeta}  \vhalf \right]
\end{align}
The clear difference to (\ref{eq:A_model}) with this choice of Lagrangian submanifold  is that there is no term quadratic in $\psi_*$, and that the $\pi$ fields are not auxiliary. The fermionic term is the same as in the standard B-model formulation, while the role of  $\pip{\bbeta}$ and $\pim{\bbeta}$ is to restrict  the theory to constant maps. Furthermore, since $\chi^\alpha$ doesn't feature in the part of the action independent of the antifields,  the $\phi^*_\alpha \chi^\alpha$ can be dropped from the extended action.  To put (\ref{eq:B_model_formulation2}) into the form (\ref{eq:stand_B_model}), we perform the field transformations
\begin{equation}
\pim{\bbeta '} = \pim{\bbeta} + g_{\bbeta \alpha}  \dem \phi^\alpha  \ \ \ \ \ \ \pip{\bbeta '} = \pip{\bbeta} + g_{\bbeta \alpha}  \dep \phi^\alpha \  .
\end{equation}
Again, in the curved case this should be done as a part of a canonical transformation (\ref{eq:canonical_transf}), but in the flat case this is equivalent to simply making the substitution. Now the transformations of $\psi_-^\alpha$ and $\psi_+^\bbeta$ contain the standard B-model BRST transformations, as well as $\pi$ dependent parts, but one is free to set all the $\pi$ terms to zero, which yields precisely  (\ref{eq:stand_B_model}), since the fixed point theorem ensures that the model is evaluated on constant maps.

\section{The doubled picture}
\label{sec:doubled_picture}
\setcounter{equation}{0}

In the previous section  we described two derivations of the A-model, (\ref{eq:A_model2})  and  (\ref{eq:A_model_orig2}), and a derivation of the B-model, (\ref{eq:AKSZ_B_model}), that involved a doubling of coordinates as an intermediate step.  It seems like a natural generalization to investigate AKSZ actions that treat the coordinates $X^i$ and $\tX_i$  on the same footing.  Geometrically, instead of taking the target space to be $\Pi(T \cM \oplus T^* \cM )$, the idea is to  work with two copies of the same Calabi-Yau,  $\cM$ and $\tcM$, where $\tX_i$ are coordinates on $\tcM$, and $\tpi^j$  fiber coordinates on the parity reversed cotangent bundle over $\tcM$. Throughout we will make an effort to understand the special case when the target space is Hull's doubled geometry \cite{Hull:2005hk, Dabholkar:2005ve,Hull:2006va}, meaning that $\cM$ is a $2n$-torus and $\tcM$ the T-dual $2n$-torus. 

In section \ref{subsec:general_setting} we write down the most general AKSZ action quadratic in $\pi$ and $\tpi$, and then concentrate on two reasonable generalizations of the standard A- and B-model formulations. The first is constructed using the inverses of the K\"{a}hler forms on $\cM$ and $\tcM$, and reduces to the sum of two Poisson $\sigma$-models when the mixed terms, those involving both $\pi$ and $\tpi$, are set to zero. The second case involves the K\"{a}hler forms themselves, and the AKSZ action can not be reduced to a sum of two actions that separately satisfy the master equation. In the flat space limit the master equation is naively satisfied irrespective of the global properties of the target space. However, when $\cM$ and $\tcM$ are $2n$-tori, we show that  for a subset of the second type of the aforementioned actions the master equation is only satisfied in arbitrary coordinates provided that $\cM$ and $\tcM$ are related by T-duality. In section \ref{subsection:Hulls_doubled} we discuss the topological twisting of the $N=2$ supersymmetric $\sigma$-model on Hull's doubled geometry.  In section \ref{subsection:choices_of_L_submanifold} we consider the choices of Lagrangian submanifold. The number of possibilities is large, but only two are able to yield models that localize on holomorphic or constant maps. With the others one is able to interpolate between the two types of theory by a canonical transformation, and we refer to these choices as \emph{intermediate}. Unsurprisingly, it is via the former two that one obtains the twisted models on Hull's doubled geometry, which we demonstrate in section \ref{subsec:doubled_A}.  The procedure is   straightforward from the AKSZ action constructed with the inverses of the K\"{a}hler forms. However, the AKSZ action constructed from the K\"{a}hler forms is singular when the target space is Hull's doubled geometry, and in order to obtain the twisted models one is forced to consider a limiting procedure. This has a particularly desirable feature that in the limit in which one recovers the twisted models, the actions obtained by the intermediate choices of Lagrangian submanifold are set to zero.   In section \ref{subsection:intermediate_choices} we discuss a class of solutions to the master equation that yield non-trivial theories for the intermediate choices of Lagrangian submanifold, but are not able to recover the twisted models.  They are characterized by being able to interpolate between a model evaluated on constant maps and one evaluated on holomorphic maps via a canonical transformation. From the standard BV arguments, in the absence of an anomaly, the resulting theories should be equivalent. This offers some intriguing questions as to whether this setting may offer some insight on mirror symmetry.  At this stage we are only able to speculate on some of the possibilities.

\subsection{The general setting}
\label{subsec:general_setting}

An obvious way to obtain an action that treats $X^i$ and $\tX_i$ symmetrically is to add the term  $\tomega^{\alpha \bbeta} \pi_\alpha \pi_\bbeta$ to (\ref{eq:A_model_orig2}):
\begin{equation}
\label{eq:doubled_action_1}
S = \int d^2 \mu \left[  \omega_{\alpha \bbeta }  \tpi^\alpha \tpi^\bbeta  + i \pi_\alpha \tpi^\alpha - i \pi_\bbeta \tpi^\bbeta  +\tomega^{\alpha \bbeta} \pi_\alpha \pi_\bbeta \right] \ .
\end{equation}
We will also consider the action constructed from $\omega^{-1}$  and $\tomega^{-1}$:
\begin{equation}
\label{eq:doubled_action_2}
S = \int d^2 \mu \left[  \tomega_{\alpha \bbeta }  \tpi^\alpha \tpi^\bbeta  + i \pi_\alpha \tpi^\alpha - i \pi_\bbeta \tpi^\bbeta  +\omega^{\alpha \bbeta} \pi_\alpha \pi_\bbeta \right] \ .
\end{equation}
Both actions satisfy the master equation only if $\omega$ and $\tomega$ are constant, and in what follows we will try to understand the appropriate generalization to curved space. An important observation is that in (\ref{eq:doubled_action_2}) turning off the mixed terms just yields a sum of two Poisson $\sigma$-models, and the master equation is then obeyed in curved space.  This is not true when one turns off the mixed terms in (\ref{eq:doubled_action_1}).

The most general action quadratic in $\pi$/$\tpi$ is  of the form
\begin{equation}
\label{eq:doubled_action_curved}
S = \int d^2 \mu \left[  \frac{1}{2} P_{i j  }(X, \tX)  \tpi^i \tpi^j  + W^{ i}_{ \ j}(X, \tX) \pi_i \tpi^j +\frac{1}{2} Q^{ i j}(X, \tX) \pi_i \pi_j \right] \  ,
\end{equation}
where all of the tensors $P$, $W$, and $Q$ depend on both $X$ and $\tX$, and the conditions obtained by imposing that $S$ satisfies the master equation (\ref{eq:master_equation}) are,
\begin{align}
\label{eq:gen_master_eq_conditions}
& P_{[ij}^{ \ \  , k} P_{|k| m]} + P_{[ ij,|k|}  W^{ k }_{ \ m ]} = 0 \ \ \ \ \ , \ \ \ \ \ Q^{[ij}_{ \ \ ,k} Q^{|k| m ]} - Q^{[ ij, |k| } W^{m]}_{ \ k} = 0  \ ,\\ \nonumber
& \frac{1}{2} P_{ij,k} Q^{km} + W^m_{ \ [ i, |k|} W^k_{ \ j]} + W^{m \  ,k}_{ \ [i} P_{|k| j ]}  - \frac{1}{2} P_{ij}^{ \ \ ,k} W^m_{ \ k} = 0  \ , \\ \nonumber
& \frac{1}{2} Q^{ij, k} P_{km} + W^{ [ i \ , |k|}_{ \ m} W^{j]}_{\ k} - W^{[i}_{ \ m ,k} Q^{|k| j } + \frac{1}{2} Q^{ij}_{ \ \ ,k} W^k_{ \ m} = 0 \ ,
\end{align}
where the comma notation is shorthand for:
\begin{align}
A(X, \tX)_{,k} := \frac{\partial A(X, \tX)}{\partial X^k} \ \ \ \mathrm{and} \ \ \  A(X, \tX)^{,k} := \frac{\partial A(X, \tX)}{\partial \tX_k} \ .
\end{align}
To generalize (\ref{eq:doubled_action_1}) to curved space we make the ansatz
\begin{equation}
\label{eq:case_omega}
P_{ij} = P_{ij}(X) \equiv \omega_{ij}( X)   \ \ \ \ \ \ Q_{ij} = Q^{ij}( \tX) \equiv \tomega^{ij} (\tX) \ 
\end{equation}
and for (\ref{eq:doubled_action_2})
\begin{equation}
\label{eq:case_omega_inverse}
P_{ij} = P_{ij}(\tX) \equiv \tomega_{ij}( \tX)   \ \ \ \ \ \ Q_{ij} = Q^{ij}( X) \equiv \omega^{ij} ( X ) \ .
\end{equation}

Let us first consider (\ref{eq:case_omega}), and suppose that we know the K\"{a}hler form on $\cM$, $\omega_{\alpha \bbeta}$, explicitly, and expand it in normal coordinates\footnote{For the use of normal coordinates on a K\"{a}hler manifold see \cite{Higashijima:2002fq}.}:
\begin{align}
\label{eq:riemann_normal}
 \omega_{\alpha \bbeta} & =  i \left( \delta_{\alpha \bbeta} + R_{\alpha \bbeta \gamma \btau} |_{X_{0}}  X^{\gamma} X^{\btau} + \cdots \right) \ .
\end{align}
We also expand $\tomega$ as
\begin{align}
\label{eq:riemann_normal2}
\tomega^{\alpha \bbeta} &=  \pm i \left( \delta^{\alpha \bbeta} + \tR^{\alpha \bbeta \gamma \btau} |_{\tX_{0}} \tX_{\gamma} \tX_{\btau} + \cdots \right) \ ,
\end{align}
but do not a priori assume a relation between $\tR$ and the Riemann tensor on $\tcM$, instead letting the master equation determine this, as well as the appropriate form of $W$. The choice of sign in (\ref{eq:riemann_normal2}) turns out to be important.  This is, of course, relative to the signs of the other terms, but the first term in the expansion of $W$ is already determined, $W^\alpha_{ \ \beta} = i \delta^\alpha_\beta$, $W^\balpha_{\bbeta} = -i \delta^\balpha_\bbeta$, in order to agree with the flat space limit (\ref{eq:doubled_action_1}).  

For  the $-$ sign case the master equation (\ref{eq:gen_master_eq_conditions})  is solved, to the first order in the expansion in $R$/$\tR$,  provided that
\begin{equation}
\label{eq:W_minus_sign_case}
W^\nu_{ \ \alpha} = i \left( \delta^\nu_\alpha + R_{\alpha \bbeta \gamma}^{ \ \ \ \  \nu} X^{\gamma} X^{\bbeta} + \tR^{\nu \bbeta \gamma}_{ \ \ \ \ \alpha} \tX_{\gamma} \tX_{\bbeta} + R_{\alpha \bbeta}^{ \ \  \ \btau \nu } X^\bbeta \tX_{\btau} + \cdots \right) \ ,
\end{equation}
$W^\bnu_{ \ \balpha}$ is the complex conjugate, and that 
\begin{equation}
\label{eq:Riemann_identification0}
 R_{ \gamma \btau}^{ \ \ \ \bbeta \alpha}|_{X_{0}} = \tR^{\alpha \bbeta}_{ \ \ \ \btau \gamma }|_{\tX_{0}}  \  .
\end{equation}
Thus, $\omega_{\alpha \bbeta}$ and $\tomega^{\alpha \bbeta}$ are expansions of the same K\"{a}hler form, and $W$ can be understood as its complexification. A less desirable feature of this solution is that in the flat space limit the number of degrees of freedom is halved, as we will explain shortly. 

For the $+$ sign case in (\ref{eq:riemann_normal2}) the master equation is solved provided that $W$ has the expansion
\begin{equation}
\label{eq:W_expansion}
W^\nu_{ \ \alpha} = i \left( \delta^\nu_\alpha + R_{\alpha \bbeta \gamma}^{ \ \ \ \  \nu} X^{\gamma} X^{\bbeta} + \tR^{\nu \bbeta \gamma}_{ \ \ \ \ \alpha} \tX_{\gamma} \tX_{\bbeta} + R_{\alpha \bbeta}^{ \ \  \ \btau \nu } X^\bbeta \tX_{\btau} + \cdots \right) \ ,
\end{equation}
$W^\bnu_{ \ \balpha}$ is the complex conjugate and
\begin{equation}
\label{eq:Riemann_identification}
 R_{ \gamma \btau}^{ \ \ \ \bbeta \alpha}|_{X_{0}} = - \tR^{\alpha \bbeta}_{ \ \ \ \btau \gamma }|_{\tX_{0}}  \  .
\end{equation}
Now,  the condition (\ref{eq:Riemann_identification}) implies that to the lowest order in $R$ the expansion of $\tomega$ corresponds to the inverse of the K\"{a}hler form. So we are actually constructing a bi-vector on $\tcM$, rather than a 2-form. Thus the solution no longer treats objects on $\cM$ and $\tcM$ on the same footing in the AKSZ action, but it avoids the problem with the reduction of degrees of freedom. 

For the (\ref{eq:case_omega_inverse}) case, we first expand $\omega^{-1}$ as
\begin{align}
\label{eq:riemann_normal3}
 \omega^{\alpha \bbeta} & =  i \left( \delta^{\alpha \bbeta} - R_{\alpha \bbeta}^{ \ \  \gamma \btau} |_{X_{0}}  \tX_{\gamma} \tX_{\btau} + \cdots \right) \ ,
\end{align}
and the master equation requires that $W$ is a complexification of the inverse K\"{a}hler form, in the same sense as (\ref{eq:W_minus_sign_case}) is a complexification of the K\"{a}hler form. The  expansion of $\tomega_{\alpha \bbeta}$ is again determined by the master equation, and as expected it corresponds to the expansion of the inverse K\"{a}hler form. 

In the flat space limit the master equation is satisfied simply because nothing in (\ref{eq:doubled_action_1})/(\ref{eq:doubled_action_2})  depends on $X$ or $\tX$.  However, for the flat space limit of the $-$ sign solution in (\ref{eq:riemann_normal2}), one obtains the condition the master equation is satisfied in curvilinear coordinates only when $\cM$ and $\tcM$ are related by T-duality.   We demonstrate this for the simplest case when $\cM$ and $\tcM$ are 2-tori, with \emph{real} coordinates $X^1$, $X^2$,$\tX_1$, $\tX_2$, which run from $-\frac{1}{2}$ to $\frac{1}{2}$, and
\begin{equation}
\label{eq:forms01}
\omega_{1 2} = m  \ \ \ \  \ , \ \ \ \ \ \tomega^{1 2 } =  \tm \ .
\end{equation}
The forms 
\begin{align}
\omega^{'}_{12} &  = \frac{4 \sin^2(\frac{1}{2} k \pi) }{k^2 \pi^2} \frac{m}{(1-4 \sin^2(\frac{k \pi}{2}) (X^1)^2 )^{\frac{1}{2}} (1- 4 \sin^2(\frac{k \pi}{2} ) (X^2)^2 )^{\frac{1}{2}} } \\ \nonumber
\tomega^{' 12} & = \frac{4 \sin^2(\frac{1}{2} k \pi) }{k^2 \pi^2} \frac{ \tm}{(1-4 \sin^2(\frac{k \pi}{2}) (\tX_1)^2 )^{\frac{1}{2}} (1- 4 \sin^2(\frac{k \pi}{2} ) (\tX_2)^2 )^{\frac{1}{2}} } \ 
\end{align}
are related to (\ref{eq:forms01}) by the diffeomorphism
\begin{align}
\label{eq:diff_of_2_torus}
X^i \rightarrow \frac{1}{2 \sin( \frac{k \pi}{2}) } \sin ( k \pi  X^i )   \ \ \  , \  \ \   \tX_i  \rightarrow \frac{1}{2 \sin( \frac{k \pi}{2}) } \sin ( k \pi  \tX_i )  \ ,
\end{align}
parameterized by $0 < k < 1$. Here $i = \{ 1, 2 \} $, and the normalization is determined by requiring the  $-\frac{1}{2}$ to $\frac{1}{2}$ range of coordinates to be preserved. The $k \rightarrow 0$ limit  takes us back to (\ref{eq:forms01}), but for the master equation to be satisfied for non-zero $k$, one needs  to take $W^{'}$ to be the 2-form on  $\cM \times \tcM$ whose only non-zero components are
\begin{align}
W^{' i }_{ \ i} &  = \frac{4 \sin^2(\frac{1}{2} k \pi) }{k^2 \pi^2} \frac{1}{(1-4 \sin^2(\frac{k \pi}{2}) (X^1)^2 )^{\frac{1}{2}} (1- 4 \sin^2(\frac{k \pi}{2} ) (\tX_1)^2 )^{\frac{1}{2}} }  \ ,
\end{align}
and it is also necessary that 
\begin{equation}
\label{eq:inverse_metric_relation}
m = \frac{1}{\tm} \ .
\end{equation}
The latter requirement can be understood by looking at the last two lines in (\ref{eq:gen_master_eq_conditions}). Only the last term in each equation is zero by itself, and since $W^{'}$ doesn't depend on $m$ or $\tm$, this dependence must also drop from the first term in each of these equations.   The equations on the first line of (\ref{eq:gen_master_eq_conditions}) are automatically satisfied since $\cM$/$\tcM$ are two dimensional.  But it's easy to see that this is also automatic if we extend the above diffeomorphisms to a $d=2 n$ dimensional torus, since the first term vanishes, and the contracted index in the second term has to be different to $i$ and $j$, but $\omega$ only depends on the $i$ and $j$ coordinates.  We stress that the master equation does not impose the T-duality condition when considering (\ref{eq:doubled_action_2}) in curvilinear coordinates. 

A feature of the solution we just described is that it is degenerate. Namely, expressing it in flat complex coordinates, so $\omega_{1 \tone} = \tomega_{1 \tone} = m $, the action (\ref{eq:doubled_action_1}) can be written as
\begin{equation}
\label{eq:degenerate_action}
\int d^2 \mu  \left[ \vhalf -i m ( i \tpi^1 - \frac{i}{m} \pi_\bone) ( i \tpi^\bone - \frac{i}{m} \pi_1 ) \right] \ .
\end{equation}
Thus the number of independent fields is halved, and we are essentially back to (\ref{eq:Poisson_action}). It turns out that it's necessary to perform a limiting procedure in order to obtain a doubled topological model. The halving of degrees of freedom does not occur when $\omega_{1 \tone} = - \tomega_{1 \tone}$, so in particular $+$ sign solution in (\ref{eq:riemann_normal2}) does not have this feature.  Furthermore, the curved space Lagrangian for the $-$ sign solution can also not be written as a square of something at arbitrary points in field space. This can be understood from the action in its general form (\ref{eq:doubled_action_curved}), which can be written as a square provided that
\begin{equation}
\tomega^{\alpha \bbeta} = W^{\bbeta}_{ \ \bnu} \omega^{\eta \bnu} W^{\alpha}_{ \ \eta} \ .
\end{equation}
Writing out the expansions of the forms one can see that this does not hold in general.

\subsection{Twisting in Hull's doubled formalism}
\label{subsection:Hulls_doubled}

In this section we derive the twisted action from a supersymmetric $\sigma$- model on Hull's doubled geometry \cite{Hull:2005hk, Dabholkar:2005ve,Hull:2006va}. Let us first consider the setting when $\cM$ and $\tcM$ are 2-tori, which is easily generalized. The bosonic action is 
\begin{align}
\label{eq:hull_action1}
S_{\bos} = & \int d^2 z \left[ \vhalf  (R_1)^2 \depp x^1 \demm x^1  +  (R_2)^2 \depp x^2 \demm x^2 + (R_1)^{-2} \depp \txx_1 \demm \txx_1 \right.  \\ \nonumber
 & + (R_2)^{-2} \depp \txx_2 \demm \txx_2  +2 \depp x^1 \demm \txx_1 - 2 \demm x^1 \depp \txx_1  \\ \nonumber 
 & \left.  +2 \depp x^2 \demm \txx_2 - 2 \demm x^2 \depp \txx_2 \vhalf \right] \ .
\end{align}
where $x^1$, $x^2$, and $\txx_1$,  $\txx_2$, are all real coordinates ranging from $0$ to $1$, with the endpoints identified. The metric components are:
\begin{equation}
g_{11} = (R_1)^2 \ \ \ , \ \ \ g_{22} = (R_2)^2 \ \ \ , \ \ \  \tg^{11} = (R_1)^{-2} \ \ \ ,  \ \ \ \tg^{22} = (R_2)^{-2} \ .
\end{equation}
The reason for the doubled plus/minus notation is that we'll shortly introduce worldsheet spinors, which carry single indices.  The topological term in the second line of (\ref{eq:hull_action1}) is crucial for establishing quantum equivalence with the standard bosonic string $\sigma$-model \cite{Berman:2007vi}, which is derived by imposing the constraints
\begin{align}
\depp (R_i X^i + \frac{1}{R_i} \tX_i ) = 0 \ \ \ \ \ \  \demm(R_i X^i - \frac{1}{R_i} \tX_i )  = 0 \ ,
\end{align}
where the repeated indices are \emph{not} summed over. For details of how the doubled action is quantized the reader is referred to the cited literature. We are also not going to be careful about overall normalization conventions for the action.

The complex structure on $\tcM$ is parameterized by $\frac{R_1}{R_2}$, 
\begin{equation}
\phi^1 = \sqrt{ \frac{R_1}{R_2} } x^1 + i \sqrt{ \frac{R_2}{R_1} } x^2 \  \ \ , \ \ \ \phi^{\bone} = \sqrt{ \frac{R_1}{R_2} } x^1 - i \sqrt{ \frac{R_2}{R_1} } x^2 \ , 
\end{equation}
and the K\"{a}hler structure by $R_1 R_2 $  , 
\begin{equation}
\omega_{1 \bone}= i g_{1 \bone} = i  R_1 R_2 \ .
\end{equation}
The quantities on  $\tcM$ are related to the above by T-duality:
\begin{equation}
\label{eq:tcoords}
\tphi_1 = \sqrt{ \frac{R_2}{R_1} } \txx_1 - i \sqrt{ \frac{R_1}{R_2} } \txx^2 \ \ \ , \ \ \  \tphi_{\bone} = \sqrt{ \frac{R_2}{R_1} } \txx_1 + i \sqrt{ \frac{R_1}{R_2} } \txx^2 \  \ \  , \ \ \  \omega_{1 \bone}= i \frac{1}{ R_1 R_2} \ .
\end{equation}
In complex coordinates  (\ref{eq:hull_action1}) reads
\begin{align}
S_\bos  = & \frac{1}{2}  \int d^2 z \left[ \vhalf R_1 R_2 ( \depp \phi^1 \demm \phi^{\bone} + \depp \phi^{\bone} \demm \phi^1 )  \right. \\ \nonumber
 & + \frac{1}{ R_1 R_2} ( \depp \tphi_1 \demm \tphi_{\bone} + \depp \tphi_{\bone} \demm \tphi_1 )  +2 \depp \phi^1 \depp \tphi_1 -2 \demm \phi^1 \depp \tphi_1  \\ \nonumber
 & \left. + 2\depp \phi^{\bone} \depp \tphi_{\bone} - 2 \demm \phi^{\bone} \depp \tphi_{\bone}  \vhalf \right]  \ ,
\end{align}
and its supersymmetrization is:
\begin{align}
\label{eq:hull_susy}
S  = & S_{\bos} + \int d^2 z \left[ \vhalf  R_1 R_2 \left( \psi_-^1 \depp \psi_-^{\bone}   + \psi_-^{\bone} \depp \psi_-^{1}   +  \psi_+^1 \demm \psi_+^{\bone} + \psi_+^{\bone} \demm \psi_+^{1}   \right) \right. \\ \nonumber 
& + \frac{1}{R_1 R_2} \left( \tpsi^+_1 \depp \tpsi^+_{\bone} + \tpsi^+_{\bone} \depp \tpsi^+_{1} + \tpsi^-_1 \demm \tpsi^-_{\bone} + \tpsi^-_{\bone} \demm \tpsi^-_{1} \right)  \\ \nonumber 
& + 2 \left( - \psi^1_- \depp \tpsi^+_1 +  \tpsi^+_1 \depp \psi^1_-  - \psi^{\bone}_- \depp \tpsi^+_{\bone} + \tpsi^+_{\bone} \depp \psi_-^{\bone}  \right. 
\\ \nonumber &  \left. \left. -  \psi^1_+ \demm \tpsi^-_{1} + \tpsi^-_1 \demm \psi^1_+   - \psi^{\bone}_+ \demm \tpsi^-_{\bone} + \tpsi^-_{\bone} \demm \psi^{\bone}_+ \right) \vhalf \right] \ .
\end{align}
The fermionic topological terms in the last two lines are necessary for establishing equivalence with the partition function of the standard supersymmetric $\sigma$-model \cite{Chowdhury:2007ba}.

Let us generalize this to a $2n$-torus, by taking $g_{\alpha \bbeta}$ to be in the diagonal form, so $g_{1 \bone} = R_1 R_2$, $g_{2 \overline{2}} = R_3 R_4$, etc., and $g_{\alpha \bbeta} = \tg_{\alpha \bbeta}$ (with the reminder that $g^{\alpha \bbeta}$ denotes the metric on $\tcM$, and $g_{\alpha \bbeta}$ is the inverse).  The A-twist  on $\cM$ is
\begin{align}
\label{eq:A_twist_on_cM}
 \psi_-^\alpha \rightarrow \psi_{--}^\alpha  \ \ \ \ \    \psi^\bbeta_+ \rightarrow \psi^\bbeta_{++}  \ \ \ \ \  \psi_+^\alpha \rightarrow \chi^\alpha     \ \ \ \ \    \psi_-^\bbeta \rightarrow \chi^\bbeta \ .
\end{align}
Of course, we could equivalently choose the complex conjugate of this. However, once the choice for $\cM$ has been made, there is no longer any freedom in the choice for $\tcM$. Namely, for the fermionic topological term in (\ref{eq:hull_susy}) to be a scalar after twisting, (\ref{eq:A_twist_on_cM}) determines the twist on $\tcM$ to be:
\begin{align}
\tpsi^-_\alpha \rightarrow \tpsi^{--}_\alpha \ \ \ \ \  \tpsi^+_\bbeta \rightarrow \tpsi^{++}_\bbeta \ \ \ \  \  \tpsi^+_\alpha \rightarrow \tchi_\alpha    \ \ \ \ \   \tpsi^-_\bbeta \rightarrow \tchi_\bbeta  \ .
\end{align}
With similar reasoning, up to overall complex conjugation, the B-twist is:
\begin{align}
& \psi_+^\bbeta \rightarrow \chi^\bbeta \ \ \ \ \   \psi_-^\bbeta \rightarrow \chi^\bbeta \ \ \ \ \   \psi_-^\alpha \rightarrow \psi^\alpha_{--} \ \ \ \ \  \psi_+^\alpha \rightarrow \psi^\alpha_{++} \\ \nonumber
&  \tpsi^-_\alpha \rightarrow \tchi_\alpha \ \ \ \ \ \ \tpsi^+_\alpha \rightarrow \tchi_\alpha \ \ \ \ \ \tpsi^+_{\bbeta} \rightarrow \tpsi^{++}_\bbeta \ \ \ \ \ \ \tpsi^-_{\bbeta} \rightarrow \tpsi^{--}_\bbeta \ .
\end{align}

The A-model action reads
\begin{align}
\label{eq:doubA_mod_from_hull}
S  = & \int d^2 z \left[ \vhalf  g_{ \alpha \bbeta} ( \dep \phi^\alpha \dem \phi^\bbeta + \dep \phi^\bbeta \dem \phi^\alpha )  + \tg^{\alpha \bbeta}( \dep \tphi_\alpha \dem \tphi_\bbeta + \dep \tphi_\bbeta \dem \tphi_\alpha )  \right. \\ \nonumber
& +  g_{\alpha \bbeta} ( \psi^\alpha_- \dep \chi^\bbeta + \psi^\bbeta_+ \dem \chi^\alpha)  + \tg^{\alpha \bbeta} ( \tpsi_\bbeta^+ \dep \tchi_\alpha + \tpsi^-_{\alpha} \dem \tchi_\bbeta ) \\ \nonumber
&  +2 (\dep \phi^\alpha \dem \tphi_\alpha -   \dem \phi^\alpha \dep \tphi_\alpha+  \dep \phi^\bbeta \dem \tphi_\bbeta -  \dem \phi^\bbeta \dep \tphi_\bbeta )  \\ \nonumber
& \left. +   \dep ( \tchi_\alpha \psi^\alpha_- + \tpsi^+_\bbeta \chi^\bbeta ) + \dem(  \tpsi^-_\alpha \chi^\alpha + \tchi_\bbeta \tpsi_+^\bbeta) \vhalf \right] \ ,
\end{align}
where we've reverted to the old single plus/minus notation, since there are no spinors in the twisted actions.
The doubled B-model action is
\begin{align}
\label{eq:doubB_mod_from_hull}
S =  & S_{\bos} +  \int d^2 z \left[ \vhalf  g_{\alpha \bbeta} \chi^\bbeta ( \dep \psi^\alpha_- +  \dem \psi^\alpha_+) + \tg^{\alpha \bbeta} \tchi_\alpha ( \dep \tpsi^+_\bbeta + \dem \tpsi^-_\bbeta )   \right. \\ \nonumber
 &  \left.  + \dep ( \tchi_\alpha \psi^\alpha_- + \tpsi^+_\bbeta \chi^\bbeta) + \dem (\tchi_\alpha \psi^\alpha_+ + \tpsi^-_\bbeta \chi^\bbeta)  \right] \ ,
\end{align}
where  $S_{\bos} $ is the bosonic part of the action, which is the same as for the A-model.

\subsection{Choice of Lagrangian submanifold}
\label{subsection:choices_of_L_submanifold}


Expanding the action (\ref{eq:doubled_action_1}) in components, with the "preliminary" choice of Lagrangian submanifold given by (\ref{eq:superfield_exp})  and (\ref{eq:doubled_expansion}) yields:
\begin{align}
\label{eq:doubled_comp_action}
S = & \int d^2 z \left[ \vphantom{\frac{1}{2}}  \omega_{\alpha \bbeta} \tpim{\alpha} \tpip{\bbeta} + \omega_{\bbeta \alpha} \tpim{\bbeta} \tpip{\alpha} + \tomega^{\alpha \bbeta} \pip{\alpha} \pim{\bbeta} + \tomega^{\bbeta \alpha} \pip{\bbeta} \pim{\alpha}  \right. \\ \nonumber 
& -i \pim{\alpha} \tpim{\alpha} + i \pip{\alpha} \tpip{\alpha} +i \pim{\bbeta} \tpim{\bbeta} -i \pip{\bbeta} \tpip{\bbeta} \\ \nonumber
& \left. +\phi^*_\alpha( i \tchi^\alpha + \tomega^{\alpha  \bbeta} \chi_\bbeta) + \phi^*_\bbeta ( -i \tchi^\bbeta + \tomega^{\bbeta \alpha} \chi_\alpha)  + \tphi_*^\alpha ( -i \chi_\alpha + \omega_{\alpha \bbeta} \tchi^\bbeta ) + \tphi^{\bbeta}_* ( i \chi_\bbeta + \omega_{\bbeta \alpha} \tchi^\alpha ) \vphantom{\frac{1}{2}} \right] \ .
\end{align}
As in the standard A- and B-model AKSZ construction, we want to choose a more refined Lagrangian submanifold based on complex structures on $\cM$ and $\tcM$. These choices can be classified as follows. The possible field/antifeld flips are,
\begin{align}
 \apip{\alpha} & \rightarrow \psi_-^\alpha &  \apip{\bbeta} & \rightarrow \psi_-^\bbeta &  \apim{\alpha}  & \rightarrow \psi_+^\alpha  &   \apim{\bbeta} & \rightarrow \psi_+^\bbeta  \\ \nonumber
\atpip{\alpha} & \rightarrow \tpsi^-_\alpha  & \atpip{\bbeta} & \rightarrow \tpsi^-_\bbeta  &  \atpim{\alpha} & \rightarrow \tpsi^+_\alpha  & \atpim{\bbeta} & \rightarrow \tpsi^+_\bbeta   \ ,
\end{align}
accompanied by corresponding field/antifield flips for $\pi_{\pm}$ and $\tpi^{\pm}$. From these we pick two of the four components  of $\pi_*^{\pm}$ and two of the four components of $\tpi^*_{\pm}$, and leave the others unchanged.  Choices related by simultaneous complex conjugation on the worldsheet and in target space are equivalent, but this still leaves a very large number of possibilities. However, the two choices corresponding to the twist discussed in the previous subsection are special. They are: 
\begin{itemize}
\item Doubled A-model
\begin{align}
\label{eq:A_doubled_model_gauge}
\pi_{+ \alpha} & \rightarrow \psi^-_{* \alpha}   &  \pi_{+ \balpha} & \rightarrow   \pi_{+ \balpha}   
&  \pi_{ - \balpha} & \rightarrow \psi^+_{* \balpha}   & \pi_{- \alpha} &  \rightarrow \pi_{ - \alpha} \\ \nonumber
 \tpip{\alpha} & \rightarrow \tpsi^{*\alpha}_-   & \tpip{\bbeta}  & \rightarrow \tpip{\bbeta}  &   \tpim{\alpha} & \rightarrow \tpim{\alpha}  &  \tpim{\bbeta} & \rightarrow \tpsi^{*\bbeta}_+    \ ,
\end{align}
\item Doubled B-model
\begin{align}
\label{eq:B_doubled_model_gauge}
 \pip{\alpha} & \rightarrow \psi^-_{* \alpha}   & \pip{\bbeta} & \rightarrow \pip{\bbeta}  & \pim{\alpha} & \rightarrow \psi^+_{*\alpha} & \pim{\bbeta} & \rightarrow  \pim{\bbeta}   \\ \nonumber
 \tpip{\alpha} & \rightarrow \tpip{\alpha}   & \tpip{\bbeta}  & \rightarrow \tpsi^{* \bbeta}_- &   \tpim{\alpha} & \rightarrow  \tpim{\alpha} &  \tpim{\bbeta} & \rightarrow \tpsi^{*\bbeta}_+   \ .
 \end{align}
\end{itemize}
After some experimenting with canonical transformations, it is not hard to see that for (\ref{eq:A_doubled_model_gauge}) one can obtain a model that localizes on holomorphic maps, but that it's never possible to obtain a model that localizes on constant maps. The opposite is true for  (\ref{eq:B_doubled_model_gauge}), namely, it's impossible to choose a fermion such that the model localizes on holomorphic maps. The other possible gauge choices are "intermediate" between the A- and B-models, by which we mean that some gauge-fixing fermions will generate models that localizes on constant maps, while others will result in localization on holomorphic maps.

\subsection{The doubled A- and B-models}
\label{subsec:doubled_A}

In the following we will derive the twisted doubled models from section \ref{subsection:Hulls_doubled} via the ASKZ formalism. One way to do this is to start with the sum of two Poisson $\sigma$-models, namely  (\ref{eq:doubled_action_2}) with the mixed terms set to zero,  and perform the derivation of the standard A- and B-models for each, adding the appropriate topological terms at the end. But this procedure does not imply the T-dual relation between $\cM$ and $\tcM$, which therefore needs to be imposed by hand. Furthermore, it doesn't provide a construction which would eliminate all the intermediate gauge choices. This is because the mixed fermionic topological terms are not generated by  a $ \int d^2 \mu D X^k D \tX_k$ term in the AKSZ action, and an argument analogous to the one used in the twisting procedure can not be applied. 

On the other hand, starting from (\ref{eq:doubled_action_2})  the consistency of the theory in curvilinear coordinates implies that there should be a T-dual relation between the manifolds. The halving of degrees of freedom is also a desirable feature, because in order to obtain the twisted doubled models it is necessary to do so via a limiting procedure, which precisely eliminates all the intermediate choices of Lagrangian submanifold. In the what follows we consider this scenario in detail.  

The starting point is the action (\ref{eq:doubled_comp_action})  with $\omega_{\alpha \bbeta} = \tomega_{\alpha \bbeta}$. Making the choice of Lagrangian submanifold (\ref{eq:A_doubled_model_gauge}) yields:
\begin{align}
\label{eq:A_model_doubled_action}
S = &  \int  d^2 z \left[\vhalf  \omega_{\alpha \bbeta} \tpim{\alpha} \tpip{\bbeta} + \tomega^{\bbeta \alpha} \pip{\bbeta} \pim{\alpha} -i \pim{\alpha} \tpim{\alpha} -i \pip{\bbeta} \pip{\bbeta} \right.  \\ \nonumber 
&  +   \omega_{\bbeta \alpha} \tpsi^{* \bbeta}_+ \tpsi^{*\alpha}_- + \tomega^{\alpha \bbeta} \psi^-_{*\alpha} \psi^+_{* \bbeta} + i \psi^-_{*\alpha} \tpsi^{*\alpha}_- +i \psi^+_{*\bbeta} \tpsi^{*\bbeta}_+ \\ \nonumber
& \left. +\phi^*_\alpha( i \tchi^\alpha + \tomega^{\alpha  \bbeta} \chi_\bbeta) + \phi^*_\bbeta ( -i \tchi^\bbeta + \tomega^{\bbeta \alpha} \chi_\alpha)  + \tphi_*^\alpha ( -i \chi_\alpha + \omega_{\alpha \bbeta} \tchi^\bbeta ) + \tphi^{\bbeta}_* ( i \chi_\bbeta + \omega_{\bbeta \alpha} \tchi^\alpha ) \vphantom{\frac{1}{2}} \right]  \ .
\end{align}
Since the $\pi$ fields are separate from the rest, they can simply be set to zero. The rest of the action can be rewritten as
\begin{align}
S = &  \int  d^2 z \left[  \frac{1}{2}  \omega_{\bbeta \alpha} ( \tpsi^{* \bbeta}_+ + i \omega^{\nu \bbeta} \psi^-_{* \nu}) ( \tpsi^{* \alpha}_- - i \omega^{\alpha \bnu} \psi^+_{*\bnu} )  \right. \\ \nonumber 
& \left. + \frac{1}{2} \tomega^{\alpha \bbeta} ( \psi^-_{* \alpha}  - i \tomega_{\bnu \alpha} \tpsi^{* \bnu}_+ ) ( \psi^+_{* \bbeta} - i \bomega_{\bbeta \nu} \psi^{* \nu}_- )   \vhalf \right] + O(\phi^*, \tphi_*) \ .
\end{align}
To obtain the action of the doubled A-model, we make the field redefinitions
\begin{align}
& \psi^{' -}_{*\alpha} = i \omega_{\alpha \bbeta} \tpsi^{*\bbeta}_+ + \psi^-_{* \alpha}  &  \tpsi^{' * \alpha}_- =  i \tomega^{\bbeta \alpha} \psi^+_{* \bbeta} + \psi^{* \alpha}_-  \\ \nonumber 
&  \psi^{' +}_{*\bbeta} = - i \omega_{\bbeta \alpha} \tpsi^{*\alpha}_- + \psi^+_{* \bbeta}  &  \tpsi^{' * \bbeta}_+ =  -  i \tomega^{\alpha \bbeta} \psi^-_{* \alpha} + \psi^{* \bbeta}_+ \ ,
\end{align}
and also
\begin{equation}
\label{eq:ghost_transf_of_coords}
c ^\alpha =   \tomega^{\alpha \bbeta} \chi_\bbeta  + i \tchi^\alpha  \ \ \ \ \    c^\bbeta =  \tomega^{\bbeta \alpha} \chi_\alpha - i \tchi^\bbeta   \ \ \ \ \   \tc_\alpha = \omega_{\alpha \bbeta} \tchi^\bbeta  - i \chi_\alpha  \ \ \ \ \   \tc_\bbeta = \omega_{\bbeta \alpha} \tchi^\alpha + i \chi_\bbeta \ .
\end{equation}
However, at this point it is necessary to go to some point where $\omega_{\alpha \bbeta}$ is not equal to $\tomega_{\alpha \bbeta}$, because otherwise the transformations are singular, related to the observation made in the context of (\ref{eq:degenerate_action}). The action now reads
\begin{equation}
\label{eq:ref_for_appendix}
S = \frac{1}{2}  \int  d^2 z \left[ \vhalf \omega^{\alpha \bbeta} \psi^-_{* \alpha} \psi^+_{*\bbeta} + \tomega_{\bbeta \alpha} \tpsi^{* \bbeta}_+ \tpsi^{*\alpha}_-  + \phi^*_\alpha c^\alpha + \phi_\bbeta^* c^\bbeta  + \tphi^\alpha_* \tc_\alpha + \tphi^\bbeta_* \tc_\bbeta  \right]  \ ,
\end{equation}
where the primes have been dropped. The idea of the limiting procedure is therefore to violate $\omega_{\alpha \bbeta} = \tomega_{\alpha \bbeta}$ by an arbitrarily small degree, perform the field redefinition,  after which we are free to restore the equality. The canonical transformations are performed as for the standard A-model, with the obvious extension of (\ref{eq:A_model_fermion}) to the tilde fields:
\begin{equation}
\label{eq:doubled_A_model_fermion}
\Psi = - \int d^2 z \left[  \omega_{\alpha \bbeta} ( \dem \phi^\alpha \psi_+^\bbeta - \dep \phi^\bbeta \psi_-^\alpha )+ \tomega^{\alpha \bbeta}(\dem \tphi_\bbeta \tpsi^-_{\alpha} - \dep \tphi_\alpha \tpsi^+_\bbeta) \right]  \ .
\end{equation}
This gives us the BRST exact part of the action (\ref{eq:doubA_mod_from_hull}) (up to an overall factor of $\frac{i}{4}$). The topological term includes not only the usual pullbacks of the K\"{a}hler form to the worldsheet, but also the mixed topological terms.   

The choice of Lagrangian submanifold (\ref{eq:B_doubled_model_gauge}) for the doubled B-model yields:
\begin{align}
S & = \int d^2 z \left[ \vhalf \tpim{\alpha} ( \omega_{\alpha \bbeta} \tpsi_-^{* \bbeta} - i \psi^+_{*\alpha} ) + \tpip{\alpha} (\omega_{\bbeta \alpha} \tpsi^{*\bbeta}_+ + i \psi^-_{* \alpha} )  \right. \\ \nonumber
& \left. + \pim{\bbeta} (\tomega^{\alpha \bbeta} \psi^-_{* \alpha} + i \tpsi^{* \bbeta}_+ ) + \pip{\bbeta} ( \bomega^{\bbeta \alpha} \psi^+_{* \alpha} - i \tpsi^{* \bbeta}_- ) \vhalf  \right]  + O(\phi^*, \tphi_*) \ .
\end{align}
After making the field redefinitions (\ref{eq:ghost_transf_of_coords}) together with
\begin{align}
&   \psi^{' +}_{* \alpha} = \psi^+_{* \alpha} - i \omega_{\bnu \alpha } \tpsi^{* \bnu}_-  & \psi^{' -}_{* \alpha} = \psi^-_{* \alpha} + i \tomega_{\alpha \bnu} \tpsi^{*\bnu}_+ \\ \nonumber
& \tpsi^{' * \bbeta}_- = \tpsi^{* \bbeta}_- -i \omega^{\bnu \bbeta} \psi^+_{* \nu} & \tpsi^{' * \bbeta}_+ = \tpsi^{* \bbeta}_+ +i \omega^{\bbeta \nu} \psi^-_{*\nu}  \ ,
\end{align}
we obtain
\begin{align}
S & = \int d^2 z \left[ \vhalf  \omega_{\alpha \bbeta} \tpim{\alpha} \tpsi^{* \bbeta}_- + \omega_{\bbeta \alpha} \pip{\alpha} \tpsi^{* \bbeta}_+ + \tomega^{\alpha \bbeta} \pim{\bbeta} \psi^-_{* \alpha} + \tomega^{\bbeta \alpha} \pip{\bbeta} \psi^+_{* \alpha} \right] + O(\phi^*, \tphi_*)  \ .
\end{align}
From here on we mimic the B-model description given at the end of section \ref{sec:standard_constructions}, using the appropriate extension of the fermion (\ref{eq:B_special_fermion}):
\begin{align}
\Psi  =  \int d^2 z  \left[ \vhalf  \omega_{\alpha \bbeta}  ( \psi^\alpha_- \dep \phi^\bbeta + \psi_+^\alpha \dem \phi^\bbeta ) + \tomega^{\alpha \bbeta} ( \tpsi^-_\bbeta \dem \tphi_\alpha  + \tpsi^+_\bbeta \dep \tphi_\alpha ) \right]  .
\end{align}

The reason why the intermediate choices of Lagrangian submanifold are eliminated is because for them the action itself turns out to be proportional to $\omega_{\alpha \bbeta} - \tomega_{\alpha \bbeta}$, and is therefore set to zero by the limiting procedure. We will show this for a particular choice in the next subsection, but one can verify that it is true generally. 

Finally, we will argue that the doubled model calculates $|Z_{A/B}|^2$ rather than $Z_{A/B} \times Z_{A/B}$. A preliminary observation is that  the map identifying T-dual directions between $\cM$ and $\tcM$ also determines the volume form on one manifold in terms of the other (and clearly we utilize these particular volume forms when calculating the partition function). One can show that it is a holomorphic map between $\tphi_1$ and $\phi^1$, rather than an antiholomorphic one, which  respects the orientations.  In the doubled B-model observables are in correspondence with $(0,p)$-forms with values in $\wedge^q T^{(1,0)} \cM$ wedged with $(p,0)$-forms with values in $\wedge^q T^{(0,1)} \tcM$. Owing to $\tphi_\alpha$ and $\phi^\alpha$ being holomorphically related,  it follows that the partition function calculates  $|Z_B|^2$. As presented this only holds for a doubled six-torus, but it seems unlikely that the property gets destroyed once one considers a theory based a generic Calabi-Yau manifold.  Why the partition function should calculate $|Z_A|^2$ for the doubled A-model seems more difficult to understand. At this stage we can only give an argument based on the calculation of the full string partition function for a doubled torus, which is given for the bosonic string in   \cite{Berman:2007vi} and the superstring in \cite{Chowdhury:2007ba}. Namely, the instanton contribution to the partition function has a factorization $Z \times \overline Z$, and the relation to the standard formulation is obtained by taking the holomorphic square root.  A crucial role in deriving this factorization is played by the mixed topological terms. It therefore seems reasonable to  conjecture that the doubled  A-model on a pair of generic Calabi-Yau manifolds calculates $|Z_A|^2$, and that a topological term constructed from $W$ (\ref{eq:W_expansion}) should play an important role.

\subsection{Intermediate gauge choices}
\label{subsection:intermediate_choices}

The intermediate choice of Lagrangian submanifold we consider in this section consists of making the same choice as in (\ref{eq:A_doubled_model_gauge}) for $\cM$, but the complex conjugate one for $\tcM$:
\begin{align}
\label{eq:B0_doubled_model_gauge}
 \pip{\alpha} & \rightarrow \psi^-_{* \alpha}   & \pip{\bbeta} & \rightarrow \pip{\bbeta}    & \pim{\alpha} & \rightarrow \pim{\alpha}  & \pim{\bbeta} & \rightarrow  \psi^+_{* \bbeta}  \\ \nonumber
 \tpip{\alpha} & \rightarrow  \tpip{\alpha}  & \tpip{\bbeta}  & \rightarrow \tpsi^{* \bbeta}_-  &   \tpim{\alpha} & \rightarrow \tpsi^{* \alpha}_+  &  \tpim{\bbeta} & \rightarrow \tpim{\bbeta}   \ .
\end{align}
After the elimination of auxiliaries we obtain:
\begin{align}
\label{eq:B0_model_action}
S &  \int  d^2 z \left[  \vhalf  E_{\alpha \bbeta} \tpsi^{*\alpha}_+ \tpsi^{* \bbeta}_- + F^{\alpha \bbeta} \psi^-_{*\alpha} \psi^+_{* \bbeta}   \right] +O(\phi^*, \tphi_*) \ ,
\end{align} 
where
\begin{equation}
\label{eq:E_and_F}
E_{ij } := \omega_{ i j } - \tomega_{ i j }  \ \ \ \mathrm{and} \ \ \   F^{ij} :=  \tomega^{i j }  - \omega^{i j}\ .
\end{equation}
This feature that the resulting action is constructed from the $E$ and $F$ tensors is characteristic of all the intermediate choices, and is the reason why these actions are set to zero in the limiting procedure discussed previously. To obtain a non-trivial theory we need to consider the opposite setting, when $E$ and $F$ are invertible. 

The feature of (\ref{eq:B0_model_action}) we wish to illustrate is that one is able to obtain either a model evaluated on constant maps, or one evaluated on holomorphic maps, depending on the choice of gauge fixing fermion.  For the B-type case there exists a canonical transformation such that the bosonic part of the resulting action is purely complex (the reader is reminded that our conventions always result in an overall factor of $i$). This mimics the situation for the standard B-model for which the complete bosonic term is contained in the  $\delta_{\BRST}$ exact part of the action. A simple choice for the fermion that achieves this is:
\begin{align}
\Psi = &   \int  d^2 z \left[  \vhalf  - E^{\alpha \bbeta}  \tpsi^+_\alpha \dep \tphi_\bbeta + i \psi^\alpha_- \dep \tphi_\alpha - F_{\alpha \bbeta} \psi^\bbeta_+ \dem \phi^\alpha + i \tpsi^-_\bbeta \dep \phi^\bbeta  \right] \ .
\end{align}
The resulting action is given by, 
\begin{align}
\label{eq:doubled_Btype_model}
S = & \int d^2 z \left[ \vhalf i \dep \tphi_\bbeta \dem \phi^\bbeta + i \dep \tphi_\alpha \dem \phi^\alpha  \right. \\ \nonumber
&  + E^{\alpha \bbeta} \dep \tpsi^+_\alpha c_\bbeta  - i \dem \tpsi^-_\bbeta \tc^\bbeta + F_{\alpha \bbeta} \dem \psi^\bbeta_+ \tc^\alpha -i \dep \psi^\alpha_- c_\alpha \\ \nonumber
& +i E_{\alpha \bbeta} \tpsi^{* \alpha}_+ \dem \phi^\bbeta + \tpsi^{* \bbeta}_- \dep \tphi_\bbeta + i F^{\alpha \bbeta} \psi^+_{* \bbeta} \dep \tphi_\alpha + \psi^-_{*\alpha} \dem \phi^\alpha \\ \nonumber
& \left. +\phi^*_\alpha \tc^\alpha + \phi^*_\bbeta \tc^\bbeta+ \tphi^\alpha_* c_\alpha + \tphi^\bbeta_* c_\bbeta +  E_{\alpha \bbeta} \tpsi^{*\alpha}_+ \tpsi^{* \bbeta}_- + F^{\alpha \bbeta} \psi^-_{*\alpha} \psi^+_{* \bbeta}   \vhalf \right] \ ,
\end{align}
where we've performed the same coordinate transformation (\ref{eq:ghost_transf_of_coords}) as for the doubled A-model.  The bosonic term involves a contraction between fields on $\cM$ and $\tcM$, which indicates that the B-type model naturally sees a complexification of a single manifold, rather than a product. From the fixed point theorem, the transformations of the fermions imply the model is evaluated on constant maps:
\begin{equation}
\dem \phi^\alpha = \dem \phi^\bbeta = \dep \tphi_\alpha = \dep \tphi_\bbeta = 0 \ .
\end{equation}

A simple choice of gauge fixing fermion for an A-type model is,
\begin{align}
\Psi = &   \int  d^2 z \left[  \vhalf   - \tpsi^+_\alpha \dep \phi^\alpha + \tpsi^-_\bbeta \dem \phi^\bbeta + \psi_-^\alpha \dep \tphi_\alpha - \psi^\bbeta_+ \dep \tphi_\bbeta \right] \ ,
\end{align}
which yields the action:
\begin{align}
\label{eq:doubled_Atype_model}
S = &   \int  d^2 z \left[  \vhalf F_{\alpha \bbeta} \dep \phi^\alpha \dem \phi^\bbeta + E^{\alpha \bbeta} \dep \tphi_\alpha \dem \tphi_{\bbeta} \right.  \\ \nonumber
 & + \dep \tpsi^+\alpha \tchi^\alpha - \dep \tpsi^-_\bbeta \tchi^\bbeta - \dep\psi^\alpha_- \chi_\alpha + \dem \psi^\bbeta_+ \chi_\bbeta \\ \nonumber
 &  + F_{\alpha\bbeta} \tpsi^{* \alpha}_+ \dem \phi^\bbeta + F_{\alpha \bbeta} \tpsi^{*\bbeta}_- \dep \phi^\alpha + E^{\alpha \bbeta} \psi^-_{*\alpha} \dem \tphi_\bbeta + E^{\alpha \bbeta} \psi^+_{* \bbeta} \dep \tphi_\alpha  \\ \nonumber
 & \left. +\phi^*_\alpha \tc^\alpha + \phi^*_\bbeta \tc^\bbeta+ \tphi^\alpha_* c_\alpha + \tphi^\bbeta_* c_\bbeta +  E_{\alpha \bbeta} \tpsi^{*\alpha}_+ \tpsi^{* \bbeta}_- + F^{\alpha \bbeta} \psi^-_{*\alpha} \psi^+_{* \bbeta}   \vhalf \right] \ .
\end{align}
From the transformations of the fermions one can read off that the model is evaluated on holomorphic maps:
\begin{align}
\dem \phi^\bbeta = \dep \phi^\alpha= \dem \tphi_\bbeta = \dep \tphi_\alpha = 0 \ .
\end{align}
Unlike the B-type model, here the contractions of the bosonic part are only between fields on the same manifold, and in this sense the A-type model sees a product of two manifolds rather than a complexification. If there exists some physical action related to this topological model, it is reasonable to assume that the bosonic part of such an action is real. The topological terms that need to be added to (\ref{eq:doubled_Atype_model}) are:
\begin{align}
\label{eq:A_type_model_top_terms}
\int d^2 \sigma \left[ -\frac{1}{2} F_{\alpha \bbeta} ( \dep \phi^\alpha \dem \phi^\bbeta - \dep \phi^\bbeta \dem \phi^\alpha ) - \frac{1}{2} E^{\alpha \bbeta} ( \dep \tphi_\alpha \dem \tphi_\bbeta - \dep \tphi_\bbeta \dem \tphi_\alpha ) \right]  \ .
\end{align}

The central problem in understanding the significance of the intermediate choices, and in particular the feature that they are able to interpolate between A-type and B-type models via a gauge transformation, is to understand the general solution to the master equation when $E$ and $F$ are invertible. This task is certainly a very complex one, but it may gives a concrete setting for  the idea of \cite{Alexandrov:1995kv}, that the mirror map can be understood as a Wick rotation in the AKSZ formalism. The one solution considered in this paper was the + sign case in (\ref{eq:riemann_normal2}), but this is by no means exhaustive. For example, we are not able to explore the whole region of moduli space not accessible by the normal coordinate expansions,  when one or both of the manifolds are at a large curvature point in the moduli space.  One hope would be that the SYZ construction can give some insight about the existence of such solutions. Even in the flat space case, when we choose $E $ and $F$ invertible with $\omega_{\alpha \bbeta} = - \tomega_{\alpha \bbeta}$, going from the A- to the B-type formulation involves a redefinition of fields  that mixes large and small directions.  It is conceivable then that in the general setting  one should look for a solution when one of the manifolds is at small volume and high curvature, and the other at large volume and small curvature, and that  the transition between the A- and B-type models involves the SYZ map \cite{Strominger:1996it}.    A different possibility is that $\cM$ and $\tcM$ are mirror manifolds, because then the invariance under the gauge choice could be implied by the defining property of mirror symmetry, in the sense that the A- and B-type models are equivalent due to:
\begin{equation}
Z_{A} ( \cM) Z_{A} (\tcM) \equiv Z_{B} ( \cM) Z_{B} (\tcM)  \ .
\end{equation}
This is still a more naive guess than it may appear, since it's not clear how the partition functions of the intermediate models relate to the standard A- and B-model. There is  certainly a discrepancy with the standard B-model  (\ref{eq:stand_B_model}), or rather its doubled version  (\ref{eq:doubB_mod_from_hull}), because in (\ref{eq:doubled_Btype_model}) we haven't dropped half of the ghost fields, and it would seem that the observables are in correspondence with de Rham cohomology classes on $\cM \times \tcM$.  There is little help from considering the A-type side, since deciphering the role of the topological terms (\ref{eq:A_type_model_top_terms}) is some way from what we are able to do at present. We give some further remarks about this in the final section.

\section{The mathematical perspective}
\label{conclusions}

It mathematical literature it has been established for some time  that two-dimesnionsal AKSZ actions are in correspondence with Lie algebroids, while three dimensional actions are in correspondence with Courant algebroids.\footnote{For reference we have written down the definitions of these structures in Appendix \ref{lie_and_courant_algebroids}. The reader is referred to  \cite{Ikeda:2006wd, Stojevic:2008qy, Pestun:2006rj} and   \cite{Roy1, Roytenberg:2006qz}  for a more complete discussion of this correspondence, the former from the physics and and the latter from the mathematics perspective.}  The relation of the two-dimensional actions to the A- and B-models, and also to the topological string, is clearly established. In fact, the AKSZ structure underlying the A- and B-model actually corresponds to the richer structure of a Lie bi-algebroid (see Appendix \ref{lie_and_courant_algebroids}).\footnote{This is also true of more general topological models based on generalized complex structures  \cite{Pestun:2006rj}.}   It is tempting to assume that there also exists a relation between AKSZ topological membrane actions and topological M-theory.  In this paper we have argued that the twisted action on Hull's doubled geometry calculates the norm squared of the A- and B-models, rather than just the squared amplitude. What has spurred the interest in considering a relation to topological M-theory is the result from \cite{Dijkgraaf:2004te},  where  the Hitchin functional formulation of M-theory on a Calabi-Yau times a circle is shown to calculate the Wigner transform of $|Z_{A/B}|^2$ (at least this is clearly established on the B-model side).  In the following we wish to note an intriguing mathematical construction that involves a doubling of coordinates as a step in establishing a relation between Lie bi-Algebroids  and Courant algebroids, or in the AKSZ language, the relation between two- and three-dimensional topological theories.
    
One significance of the fact that the A- and B-models have a relation to Lie bi-algebroids, rather than just Lie algebroids,  is that the former can be mapped into Courant algebroids.  In the AKSZ language this map manifests itself in the fact that a three dimensional AKSZ action corresponding to a Courant algebroid which has a Lie bi-algebroid origin, is a total derivative, and reduces to a two dimensional topological theory on the boundary \cite{Ikeda:2006wd, Stojevic:2008qy}.   In the present context we are more interested in the explicit construction of this map given in \cite{Roy1}.  The crucial ingredient is the construction of the \emph{Drinfeld double} of a Lie bi-algebroid which is given by the supermanifold $T^* \Pi A$ together with a nilpotent vector field on it, where $\Pi$ denotes the parity reversal while $A$ is the vector bundle in question. In addition an isomorphism between $T^* \Pi A$ and $T^* \Pi A^*$ is established.  For the topological A- and B-models $A^*$ needs to be identified with $T^* \cM$. Now, $T^* \Pi T^* \cM$  is a manifold with twice the number of coordinates. Locally we could take  $\tX_i$ to be the coordinate tangent to $\cM$ (i.e. tangent to the even directions of $\Pi T^* \cM$),  while $\tpi_i$ would be interpreted as the coordinate tangent to the odd direction of $\Pi T^* \cM$. Globally,  $T^* \cM$ certainly can not be identified with the type of doubled manifold that arises in a T-duality invariant formulation of physics.  For the doubled torus this is obvious, since then $\tX_i$ corresponds to a compact direction, whereas on  $\Pi T^* \cM$ this direction is non-compact. It would seem that one needs to exponentiate $T^* \Pi T^* \cM$ in some way, just as one would need to exponentiate the Drinfeld double Lie algebra to obtain the group.


A further point of interest is that $T^* \Pi A$ and  $T^* \Pi A^*$ are identified via a type of Lagrange transform. If we take $T^* \Pi T^* \cM$ and $T^* \Pi T \cM$ to correspond locally with the doubled manifolds we considered in this paper, then this Lagrange transform relates the actions (\ref{eq:doubled_action_1})  and (\ref{eq:doubled_action_2}). In establish a relation between the two actions we need to take  $\omega_{\alpha \bbeta} = - \tomega_{\alpha \bbeta}$, since (\ref{eq:doubled_action_1}) is degenerate when $\omega_{\alpha \bbeta} = \tomega_{\alpha \bbeta}$. With this choice the doubled twisted model of section \ref{subsection:Hulls_doubled} can be obtained starting from (\ref{eq:doubled_action_1}), and as we have argued this calculates $|Z_{A/B}|^2$. On the other hand, these are \emph{not} the quantities calculated from (\ref{eq:doubled_action_1}), as was established in section \ref{subsection:intermediate_choices}.  Now, in \cite{Dijkgraaf:2004te, Ooguri:2004zv} it is shown that the object related to $Z_B$ by a Lagrange transform is  $(\int_\cM \Omega \wedge \overline{\Omega} )$, where $\Omega$ is the holomorphic three-form. It seems unlikely that the appearance of a Lagrange transform both on the physics and the mathematics side, in a closely related context, is an accident, and this is certainly worth investigating further. If they can be identified, the implication would be that the B-type intermediate gauge choice of section \ref{subsection:intermediate_choices}  calculates  $(\int_\cM \Omega \wedge \overline{\Omega} )^2$.

In conclusion, it would be of interest to understand the relation of the mathematics literature and the doubled formalism of physics more precisely, and to understand what kind of an extension is needed to rigorously define the doubled objects relevant for a T-duality invariant formulation of string theory. 



\vskip 1cm

{\large \bf Acknowledgments}\\ [2mm]

I am grateful to Vicente Cort\'es and Ron Reid-Edwards for useful discussions and suggestions, and to Andrew Neitzke for commenting on an earlier draft of the paper. I would also like to acknowledge the generous funding from the German Research Foundation (DFG).

\vskip 1.3cm

\appendix
\section{Notation and conventions}\label{conventions}
\setcounter{equation}{0}

\begin{itemize} 
\item We consider diffeomorphic Calabi-Yau manifolds: $\cM$ and $\tcM$. Following this notation, coordinates  and fields on $\tcM$ always carry a tilde symbol.  
\item For AKSZ actions in components antifields are denoted by $*$, while in deRham superfields $\pi_i$ is the antifield of $X^i$, and $\tpi^i$ the antifield of $\tX_i$. 
\item The BRST transformations of the topological theories are always expressed by writing down the extended Batalin-Vilkovisky action. The reader who is trying to understand the paper but is not familiar with the BV formalism, can initially read off the BRST transformations from the terms linear in the antifields worry about the AKSZ construction later. For example the BRST transformation in the topological A-model $\delta_{\BRST} \phi^i = \chi^i$,  $\delta \chi^i = 0$ is expressed in the extended action (see for example (\ref{eq:ref_for_appendix})) by the presence of the term 
\begin{equation}
\int d^2 z \phi^*_i \chi^i  \ .
\end{equation}
A term proportional to $\chi^*_i$ is missing precisely because the transformation of $\chi$ is trivial.
\end{itemize}

\section{Definitions of Lie and Courant algebroid structures}\label{lie_and_courant_algebroids}
\setcounter{equation}{0}

A Lie algebroid  \cite{Mackenzie} $L$   is a vector bundle on $\cM$ together with a Lie bracket $[ \cdot , \cdot ]$ that acts on sections of $L$, and an anchor map $a: C^{\infty} (L) \rightarrow C^{\infty} (T \cM)$ that satisfies:
\begin{align}
\label{eq:Lie_algebroid}
& a ( [ X,Y]) = [a(X) , a(Y) ] \\ \nonumber
& [X, fY ] = f [ X, Y] + (a (X) f) Y \ \ \ \forall X, Y \in C^{\infty} (L), f \in C^{\infty} (\cM) \ .
\end{align} 
There exists a natural exterior derivative $d_L: C^\infty(\Lambda^k L^*) \rightarrow C^\infty(\Lambda^{k+1} L^*)$, where $L^*$ is the dual space to $L$  that obeys $(d_L)^2=0$. If $d_L$ obeys the Leibnitz rule  with respect to the Lie algebroid bracket on $L^*$,  then $(L, L^*)$ is said to have the structure of a \emph{Lie bi-algebroid}. 

A Courant algebroid \cite{LWX, Roy1}  is a vector bundle $E$ with a bilinear form $\langle , \rangle$, a bracket $[ , ]$, and an anchor map $\pi: E \rightarrow T \cM$ obeying
\begin{align}
\label{eq:def_Courant_alg}
& \pi ( [ X,Y]) = [ \pi (X) , \pi (Y) ]  \ \ \ \forall X, Y \in C^{\infty} (E), \\ \nonumber
& [X, fY ] = f [ X, Y] + (a (X) f) Y + (\pi(X) f) Y - \langle X, Y \rangle \cD f \ \ \ \forall X, Y \in C^{\infty} (E), f \in C^{\infty} (\cM) \\ \nonumber
& \langle \cD f, \cD g \rangle = 0 \ \ \ \forall f, g \in C^{\infty} (\cM)  \\ \nonumber 
& \pi (X) \langle Y, Z \rangle - \langle [X, Y ] + \cD \langle X, Y \rangle, Z \rangle + \langle Y, [X, Z] + \cD \langle X, Z \rangle \rangle \ \ \ \forall X, Y, Z \in C^\infty (E)  \\ \nonumber 
& [X, [Y, Z]] + [Y, [Z, X]] + [Z, [X,Y]] = \frac{1}{3} \cD (  \mathrm{Nij} (X, Y, Z) ) \ \ \  \forall X, Y, Z  \in C^{\infty} (E)  \ ,
\end{align} 
where
\begin{equation}
\mathrm{Nij} (X, Y, Z) :  =  \langle [X, Y], Z \rangle +  \langle [Y, Z], X \rangle  +  \langle [Z, X], Y \rangle \ ,
\end{equation}
and $\cD$ is a map $C^\infty(\cM) \rightarrow C^\infty(E) $ defined by the property $\langle \cD f, X \rangle = \frac{1}{2} \pi (X) f$  \ $\forall f \in C^\infty(\cM),  X \in C^\infty (E)$.


\end{document}